\documentclass[lettersize,journal]{IEEEtran}
\usepackage{amsmath,amsfonts}
\usepackage{algorithmic}
\usepackage{algorithm}
\usepackage{array}
\usepackage[caption=false,font=normalsize,labelfont=small,textfont=small]{subfig}
\usepackage{textcomp}
\usepackage{stfloats}
\usepackage{url}
\usepackage{verbatim}
\usepackage{graphicx}
\usepackage{cite}
\usepackage{url}
\usepackage{color}
\usepackage{multirow}

\newcommand{\edita}[1]{\textcolor{black}{#1}}
\newcommand{\editb}[1]{\textcolor{black}{#1}}
\newcommand{\editc}[1]{\textcolor{black}{#1}}
\begin{document}

\title{RDP-Net: Region Detail Preserving Network for Change Detection}


\author{Hongjia Chen,
Fangling Pu,~\IEEEmembership{Member,~IEEE},
Rui~Yang,
Rui~Tang,
Xin~Xu
\thanks{© 2022 IEEE.  Personal use of this material is permitted.  Permission from IEEE must be obtained for all other uses, in any current or future media, including reprinting/republishing this material for advertising or promotional purposes, creating new collective works, for resale or redistribution to servers or lists, or reuse of any copyrighted component of this work in other works.
(Digital Object Identifier 10.1109/TGRS.2022.3227098)}
\thanks{
This work was supported in part by National Natural Science Foundation of China (Grant: 62071336) and in part by the National Key Research and Development Program of China (Grant: 2018YFB2100503). (\textit{Corresponding author: Fangling Pu)}}
\thanks{
H. Chen, F. Pu, R. Yang, R. Tang and X. Xu are with the Collaborative Sensing Laboratory, Electronic Information School, Wuhan University, Wuhan 430079, China (e-mail: chj1997@whu.edu.cn; flpu@whu.edu.cn; ruiyang@whu.edu.cn; tangr@whu.edu.cn; xinxu@whu.edu.cn).}
}

\markboth{Journal of \LaTeX\ Class Files,~Vol.~14, No.~8, August~2021}%
{Shell \MakeLowercase{\textit{et al.}}: A Sample Article Using IEEEtran.cls for IEEE Journals}


\maketitle

\begin{abstract}
Change detection (CD) is an essential earth observation technique. It captures the dynamic information of land objects. With the rise of deep learning, convolutional neural networks (CNN) have shown great potential in CD. However, current CNN models introduce backbone architectures that lose detailed information during learning. Moreover, current CNN models are heavy in parameters, which prevents their deployment on edge devices such as UAVs.
In this work, we tackle this issue by proposing RDP-Net: a region detail preserving network for CD.
We propose an efficient training strategy that constructs the training tasks during the warmup period of CNN training and lets the CNN learn from easy to hard.
The training strategy enables CNN to learn more powerful features with fewer FLOPs and achieve better performance.
Next, we propose an effective edge loss that increases the penalty for errors on details and improves the network's attention to details such as boundary regions and small areas.
Furthermore, we provide a CNN model with a brand new backbone that achieves the state-of-the-art empirical performance in CD with only 1.70M parameters.
We hope our RDP-Net would benefit the practical CD applications on compact devices and could inspire more people to bring change detection to a new level with the efficient training strategy.
\editc{The code and models are publicly available at \url{https://github.com/Chnja/RDPNet}.}
\end{abstract}


\section{Introduction}
\IEEEPARstart{C}{hange} detection~(CD) reports the temporal dynamics of the studied area by observing it at different times~\cite{singh1989review}. In geoscience, the observation is conducted through the remote sensing technique~\cite{yang2021da2net,li2020amn}. With the information interpreted from multi-temporal satellite images, CD benefits applications such as urban planning \cite{ji2018fully}, environment monitoring \cite{masek2015role,reba2020systematic,sippel2020climate}, disaster assessment \cite{lei2019landslide,ma2016automatic,peng2020optical,doroodgar2014learning} and resource management \cite{sim2006grid}.

\edita{Traditional CD methods can be divided into two categories: Pixel-based and Object-based \cite{hussain2013change}.
The Pixel-based CD methods generate a change map by comparing multi-temporal remote sensing images pixel by pixel.
The representatives are change vector analysis (CVA)\cite{lambin1994change}, principal component analysis (PCA)\cite{celik2009unsupervised}, independent component analysis (ICA)\cite{marchesi2009ica}, etc.
Recently, some researchers use neural networks to extract deep change vectors (DCVA)\cite{saha2019unsupervised,saha2020unsupervised,saha2020building} and have made some achievements, especially in the field of unsupervised learning.
However, Pixel-based CD methods have limited capability in capturing spatial context information and complex visual features.
The Object-based methods segment the raw image into different categories and then obtain the change map via comparison \cite{castilla2008image}.
However, it is usually difficult to set suitable parameters for segmentation, limiting the performance.
Traditional CD methods tend to have high requirements for image registration, leading to a strong dependence on pre-processing approaches.}

\begin{figure}[t]
\centering
\subfloat[]{
    \includegraphics[width= 0.3\linewidth]{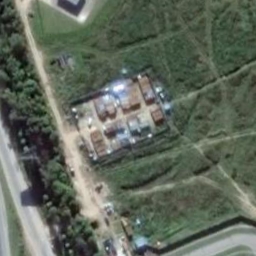}
}
\subfloat[]{
    \includegraphics[width= 0.3\linewidth]{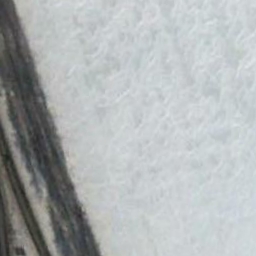}
}
\subfloat[]{
    \includegraphics[width= 0.3\linewidth]{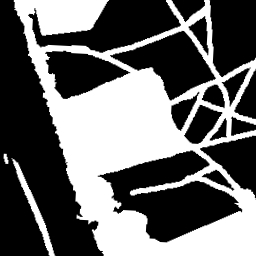}
}

\subfloat[]{
    \includegraphics[width= 0.3\linewidth]{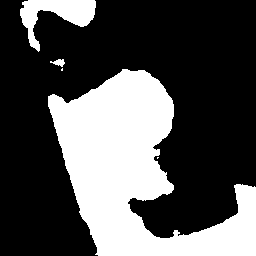}
}
\subfloat[]{
    \includegraphics[width= 0.3\linewidth]{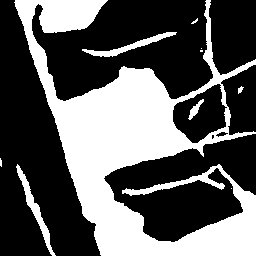}
}
\subfloat[]{
    \includegraphics[width= 0.3\linewidth]{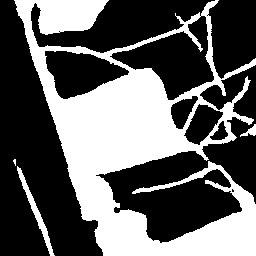}
}
\caption{Multi-temporal remote sensing images in CDD dataset.
(a) and (b) are the original images, (c) is the ground truth.
The result of (d) FC-Siam-diff, (e) SNUNet-CD, (f) our method.
The white pixels represent the changed area.}
\label{image1}
\end{figure}

Recently, convolutional neural networks (CNN) have demonstrated empirical success in plenty of geoscience applications such as seismic imaging~\cite{dxm+21}, scene classification~\cite{yang2021da2net}, texture evaluation~\cite{chen20213d}, etc.
\edita{In the field of CD, CNN has also demonstrated strong capabilities \cite{chen2019change,wu2021unsupervised}.}
Current CNN models can be mainly divided into two categories according to their backbone, U-Net-like \cite{ronneberger2015u} methods and ResNet/VGG-based \cite{he2016deep,simonyan2014very} methods.
\editb{Daudt \textit{et al.} extended the ideas of Early Fusion (EF) and Siamese (Siam) to CD \cite{daudt2018urban}, used the concept of skip connections that were used to build the U-Net and proposed FC-EF, FC-Siam-conc and FC-Siam-diff \cite{daudt2018fully}.}
These networks are usually recognized as the cornerstone of deep learning-based change detection.
\editb{Their experiments also demonstrate that directly feeding overlaying images into the network can already achieve good results, and this paper does the same.}
Peng \textit{et al.} \cite{peng2019end} and Fang \textit{et al.} \cite{fang2021snunet} proposed UNet++\_MSOF and SNUNet-CD based on Daudt's work.
They replaced U-Net with U-Net++ \cite{zhou2018unet++}, and fused the multiple side outputs of U-Net++, which achieves a better performance in the field of change detection.
Chen \textit{et al.} \cite{chen2020dasnet} proposed DASNet.
\editc{Zhang \textit{et al.} \cite{zhang2020feature} proposed FDCNN.
Zhang \textit{et al.} \cite{zhang2020deeply} proposed IFN.}
ResNet and VGG are used as backbone to extract feature maps from the multi-temporal.
A dual attention mechanism was proposed to fuse the feature map from different depths.
\editc{Li \textit{et al.} \cite{li2022remote} designed a temporal feature interaction module (TFIM) to enhance interaction between bi-temporal features and proposed TFI-GR.}
%
%
\edita{However, existing methods cannot tackle the major challenges of CD:
characterizing the dynamics of the boundary regions and small areas.}
An example is demonstrated in Fig. \ref{image1}. The changed area contains acres of buildings and a few country roads.
Although the up-to-date methods have good intuition on general changed areas, they failed to identify the boundary regions and had poor results in modeling the changes in the roads.
Only our region detail preserving network (RDP-Net) can detect the changed buildings and roads well.

{\it
Can we build a compact neural network with better modeling of details for change detection?
}

\edita{In this paper, we answer this question positively by proposing RDP-Net: a region detail preserving network for CD.
We provide an efficient training strategy, an effective edge loss and a brand new CNN backbone focusing on preserving regions’ details.}
These three factors strengthen the performance of our method in CD.

\edita{First of all}, CD in remote sensing images faces various scenes.
The difficulty of detection varies for different scenes.
\editc{Therefore, }
\edita{from the perspective of model training, network training becomes more difficult.}
\editb{Curriculum learning\cite{bengio2009curriculum} suggests that simple, basic tasks are more useful for network training.
Specifically, in this paper, an efficient training strategy is proposed to split the training dataset into different subsets by difficulty level, and let the CNN learn from easy to hard.
We argue that it is beneficial to warm up the CNN with samples from easier areas.
Next, our model would learn harder subtle areas better with previous training knowledge.}
\edita{Besides, one idea for curriculum learning is to reduce the impact of hard samples during training \cite{xu-etal-2020-curriculum,jiang2014easy}.
We believe that in CD, hard samples can help further improve the performance of the network in the last training stage, so our efficient training strategy can achieve better results in CD.}

\edita{Then}, the boundary regions and small areas are not well-processed and are still a challenging problem in remote sensing change detection.
The bad boundary performance turns out to be the bottleneck of change detection accuracy.
\edita{However, according to the analysis, the boundary regions and small areas are quite undervalued in the current CNN.}
In the existing methods, the most commonly used loss functions are CrossEntropy, Balanced Cross Entropy, Focal Loss \cite{lin2017focal}, Dice Loss \cite{milletari2016v} and IOU Loss \cite{yu2016unitbox}, \edita{which mainly aim at dealing with class imbalance problems without well considering the boundary regions and small areas.
Obviously, determining the boundaries of the changed areas is more valuable than determining the internal regions in the field of CD.
There should be a greater penalty for errors in the boundary regions.
Yet little attention is currently paid to this issue.
The Boundary Loss proposed by Hoel \textit{et al.} \cite{kervadec2019boundary} is designed for medical image segmentation, which is not suitable for change detection using remote sensing images, especially when it comes to complex scenarios.}
\edita{Thus, an Edge Loss by increasing penalty for errors on the boundary regions is proposed.
We argue that our edge loss helps the network to pay attention to the details such as boundary regions and small areas, leading to accurate detection of the details.}

\edita{At last, we argue that current CNN models have limited success in CD as their networks focus more on the global context than on the local information.
The state-of-the-art (SOTA) CNN models in CD use U-Net~\cite{ronneberger2015u} or ResNet~\cite{he2016deep} as their backbone.
While both backbone networks are designed for image recognition, they focus on the global context and discard a lot local information via pooling operations~\cite{dwivedi2020benchmarking}. If U-Net and ResNet are directly applied as the backbone, it is natural that modeling details such as boundary regions and small areas would be difficult.
In this paper, a new CNN with a brand new backbone is proposed, which is more suitable for CD tasks.}
It can sense the input images by region, focus on the local changes between the input multi-temporal images and ensure that the information is fully utilized.
\edita{It is worth mentioning that this new CNN is quite compact, with a small number of parameters, which allows it to be deployed on edge devices such as UAVs.}


The RDP-Net has achieved a better result in the field of remote sensing change detection.
The major contributions of RDP-Net can be summarized as follows:
\begin{itemize}
    \item {\it An Efficient Training Strategy:} \edita{We propose a strategy to let the network learn from easy to hard. The datasets are split by difficulty level and training tasks are constructed in the early stage of training. Experiments show that this efficient training strategy can achieve better performance.}
    \item {\it An Effective Edge Loss:} We propose an edge loss, which increases the penalty for errors on the boundary regions when calculating loss. We demonstrate that this loss improves the network's attention on the details such as boundary regions and small areas, and increases the detection accuracy.
    \item {\it A Compact Network:} We perform parameter tuning following the proposed efficient training strategy and edge loss. Experimental results indicate that our method leads us to find out a network that achieves the SOTA accuracy in remote sensing change detection with a brand new backbone. Moreover, the number of parameters in our RDP-Net is 1.70M, which enables its deployment in more democratized devices. 
\end{itemize}

This paper is organized as follows.
Section II describes the change detection method proposed in this paper.
Section III contains a series of quantitative comparisons and analyses through experiments.
Finally, the conclusion is drawn in Section IV.

\section{Methodology}


In this section, we first introduce our efficient training strategy,
splitting the dataset by difficulty level and training the CNN model from easy to hard.
Then, we propose an edge loss to improve the network's attention on the details such as boundary regions and small areas.
At last, the architecture of the proposed RDP-Net with a brand new backbone is presented.

\subsection{Efficient Training Strategy}

\begin{figure}[ht]
\centering
\subfloat[]{
    \includegraphics[width= 0.23\linewidth]{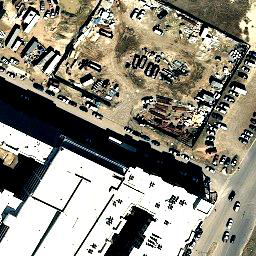}
    \label{hierarchical11}
}
\subfloat[]{
    \includegraphics[width= 0.23\linewidth]{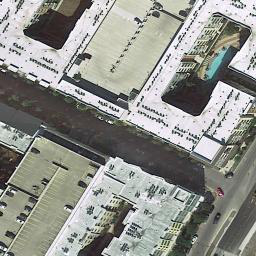}
    \label{hierarchical12}
}
\subfloat[]{
    \includegraphics[width= 0.23\linewidth]{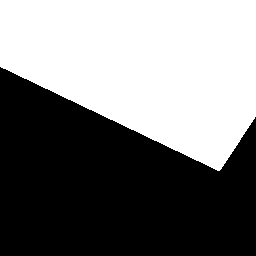}
    \label{hierarchical1l}
}
\subfloat[]{
    \includegraphics[width= 0.23\linewidth]{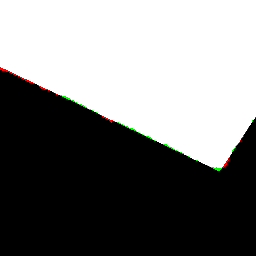}
}

\subfloat[]{
    \includegraphics[width= 0.23\linewidth]{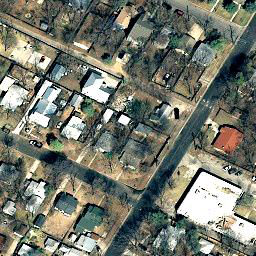}
    \label{hierarchical21}
}
\subfloat[]{
    \includegraphics[width= 0.23\linewidth]{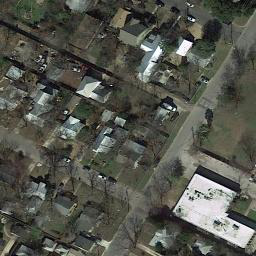}
    \label{hierarchical22}
}
\subfloat[]{
    \includegraphics[width= 0.23\linewidth]{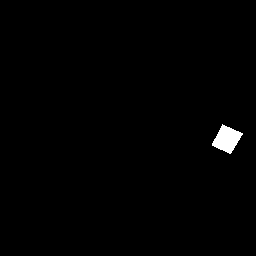}
    \label{hierarchical2l}
}
\subfloat[]{
    \includegraphics[width= 0.23\linewidth]{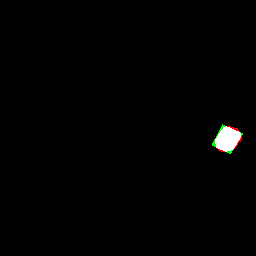}
}
\caption{(a), (b), (e) and (f) are the original input images. (c) and (g) are ground truth.
The white pixels represent the changed area.
\editc{The results of (d)(h) our RDP-Net.
The false positives and false negatives are indicated by red and green, respectively.
Other colors represent true positives.}}
\label{hierarchical}
\end{figure}

Change detection in remote sensing images faces various challenges and situations.
Some changed areas may contain thousands of pixels while some of them only take few pixels.
Meanwhile, seasonal changes are regarded as interference factors, which brings difficulties to change detection.
As shown in Fig. \ref{hierarchical}, the changed area between Fig. \ref{hierarchical}(a) and Fig. \ref{hierarchical}(b) is a building covering a large area, the changed area between Fig. \ref{hierarchical}(d) and Fig. \ref{hierarchical}(e) is much smaller and under the influence of seasonal changes (the growth of trees).
Obviously, Fig. \ref{hierarchical}(f) is more difficult to detect than Fig. \ref{hierarchical}(c).

\begin{figure}[htb]
\centering
\subfloat[]{
    \includegraphics[width= 0.3\linewidth]{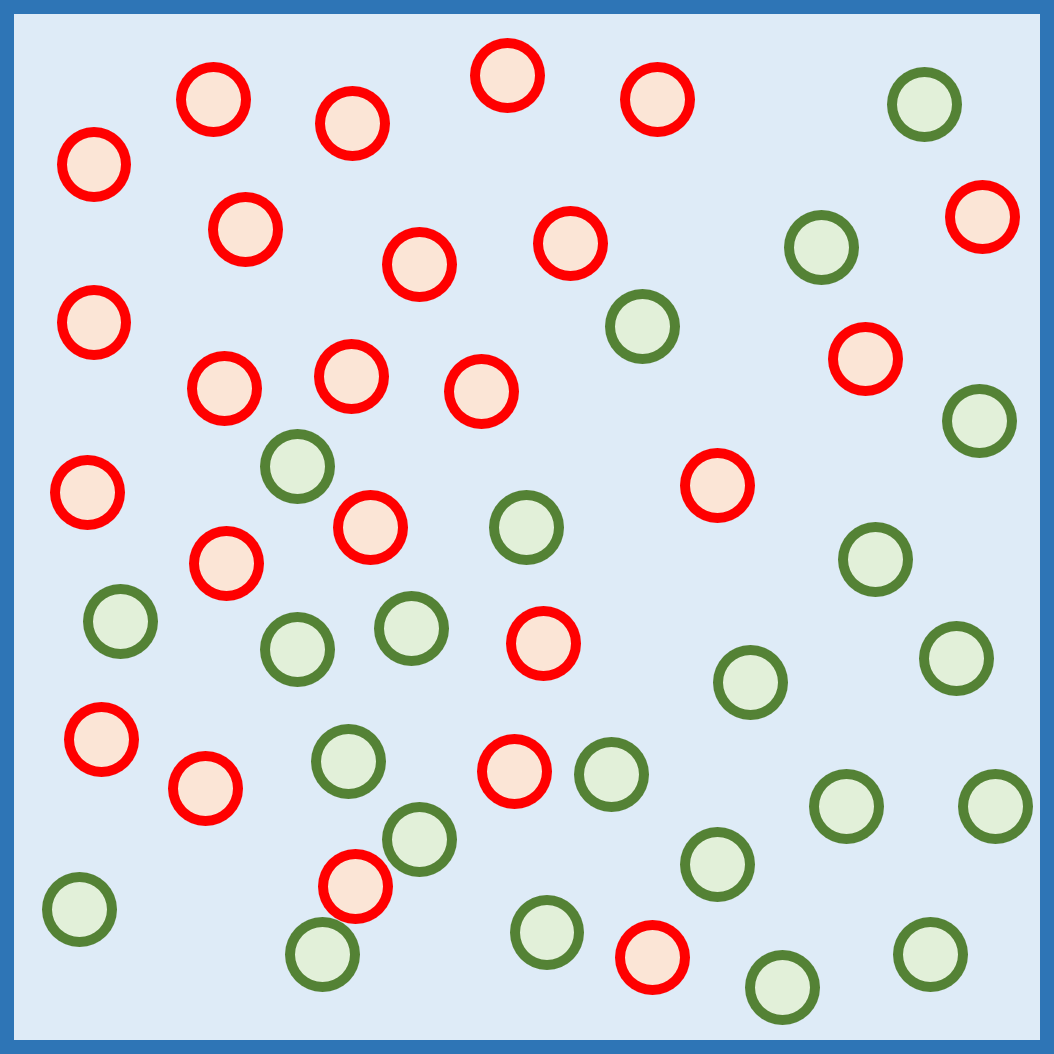}
}
\subfloat[]{
    \includegraphics[width= 0.3\linewidth]{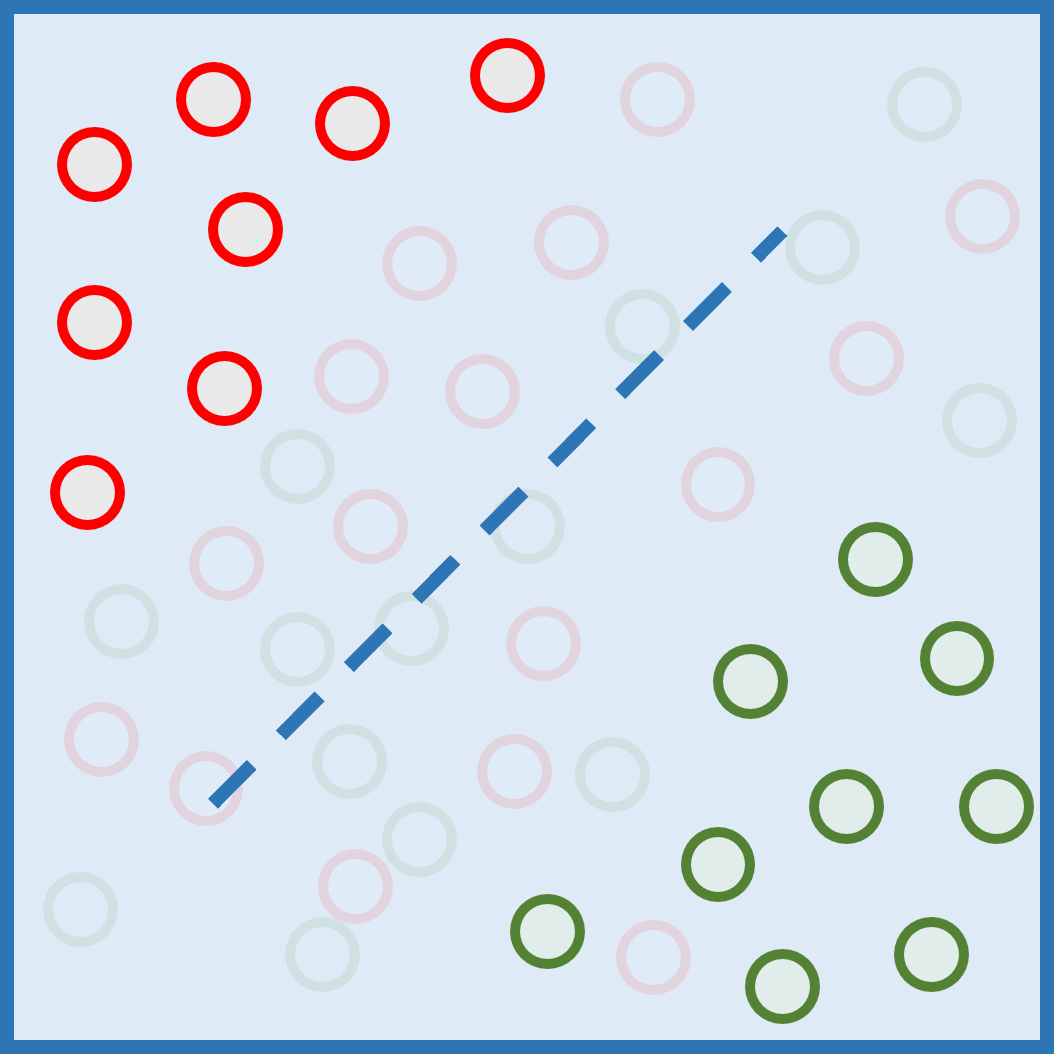}
}
\subfloat[]{
    \includegraphics[width= 0.3\linewidth]{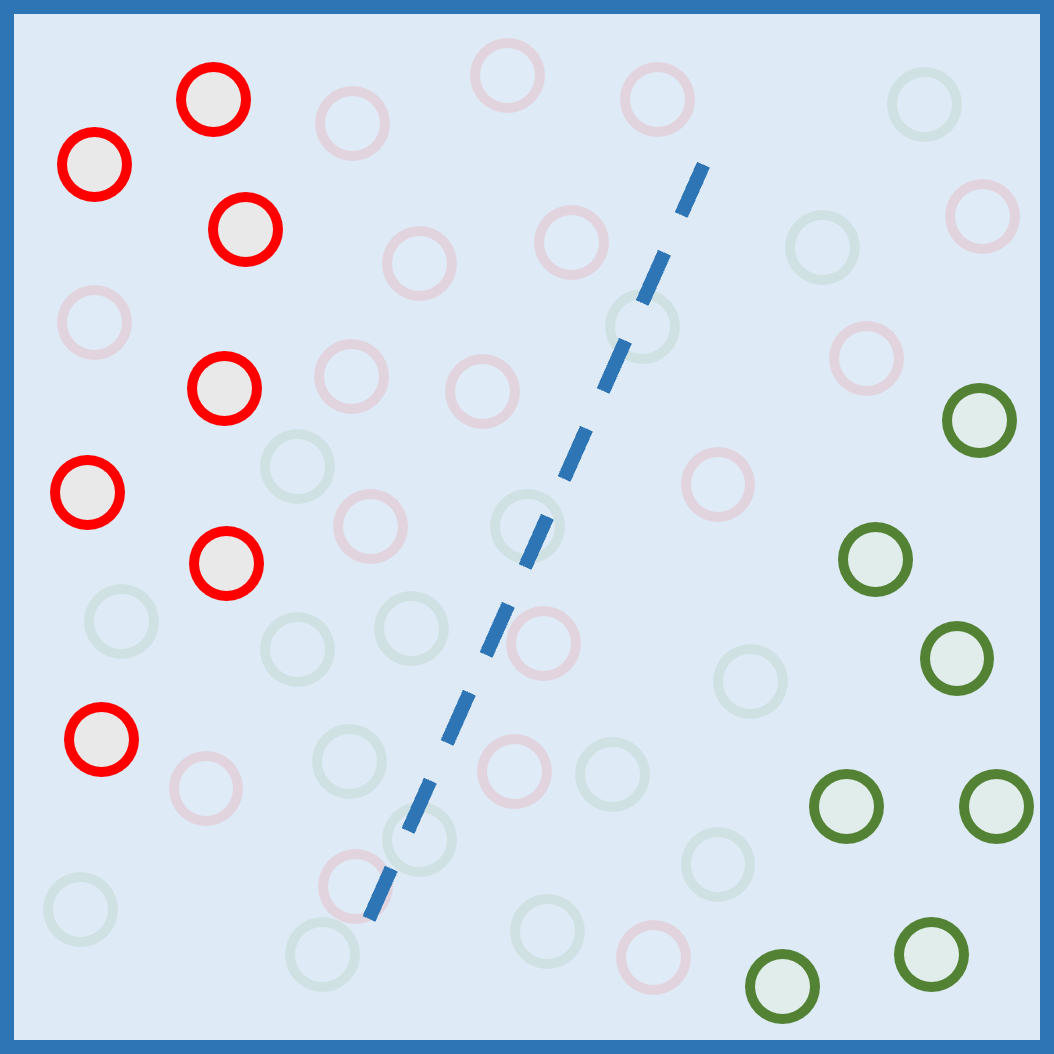}
}

\subfloat[]{
    \includegraphics[width= 0.3\linewidth]{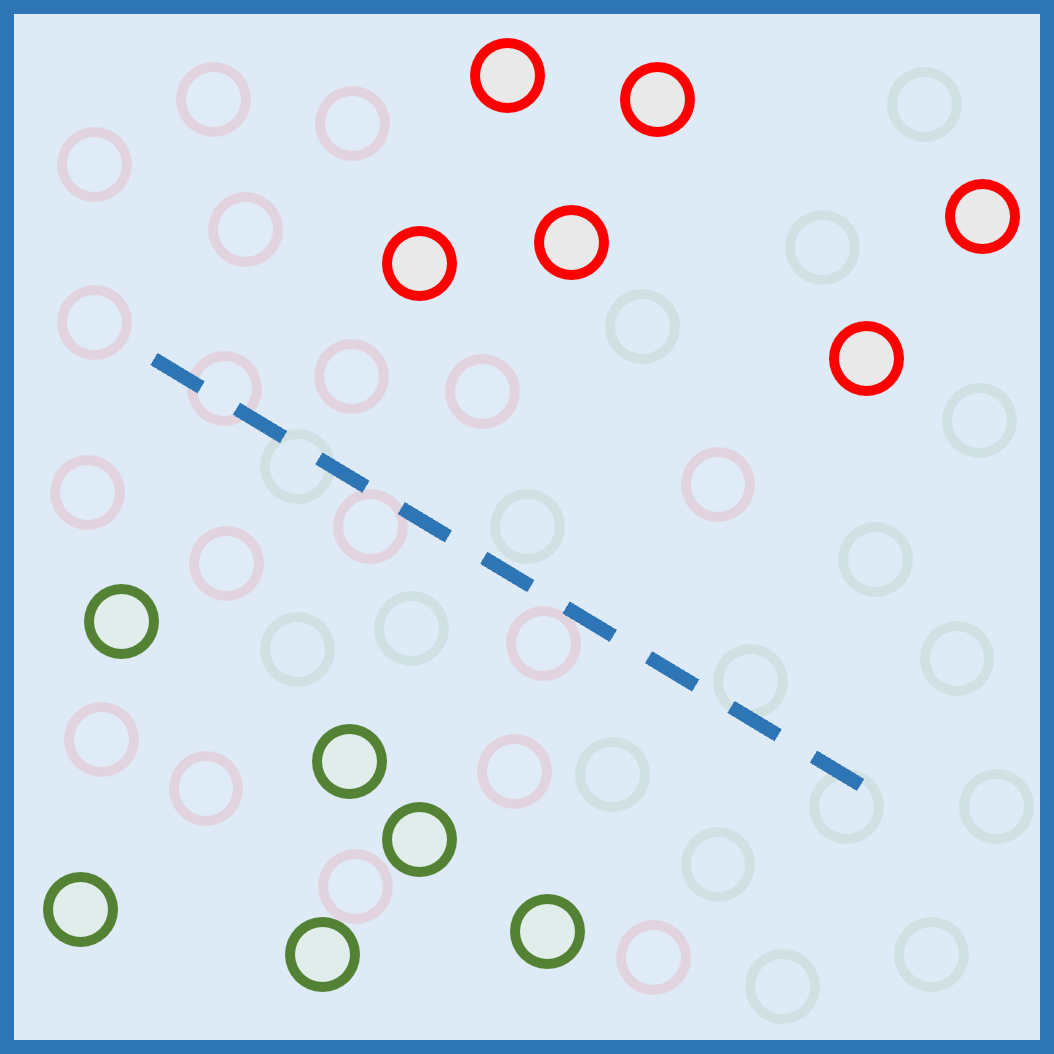}
}
\subfloat[]{
    \includegraphics[width= 0.3\linewidth]{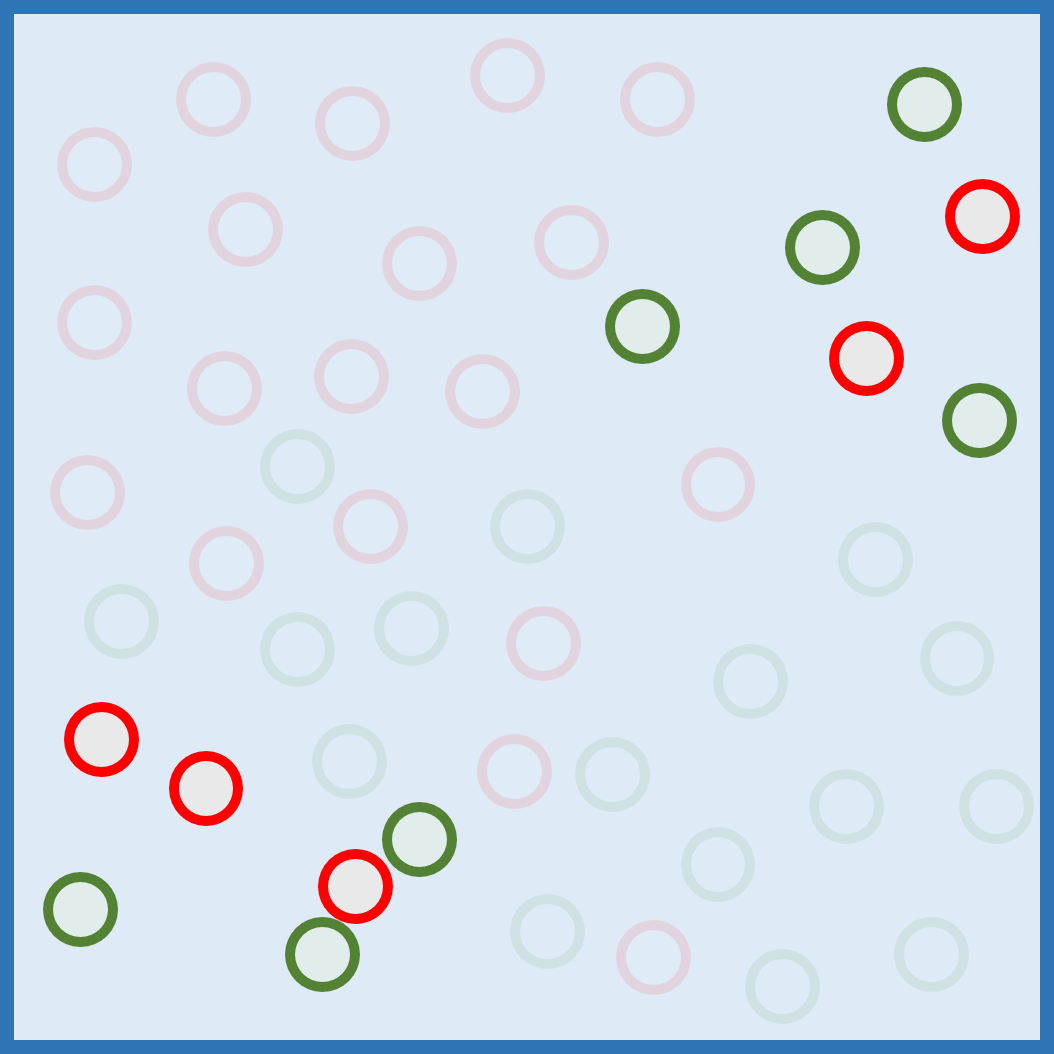}
}
\subfloat[]{
    \includegraphics[width= 0.3\linewidth]{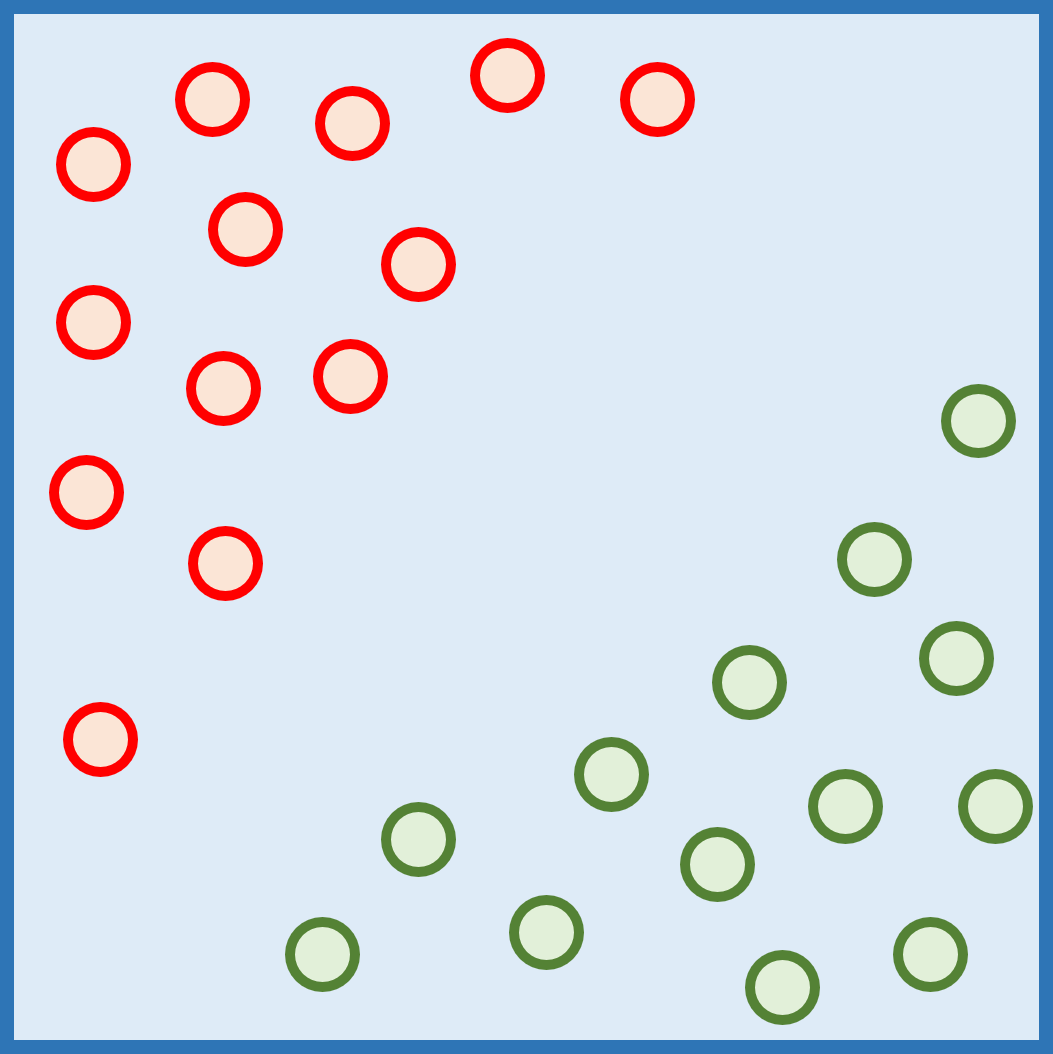}
}
\caption{An abstract image of a classification task.
(a) represents the whole dataset of the task.
(b)-(e) represent some possible batches during the training process.
(f) represents an easy subset of the task.
Different colors represent different classes.}
\label{hierarchicalshow}
\end{figure}

However, most current deep learning methods treat different samples equally.
A more adaptive strategy should be considered, since some samples may be more difficult and affect the effectiveness of learning.
As shown in Fig. \ref{hierarchicalshow}, for such a classification task, the batches used for training are randomly generated.
If all the samples are directly used for training, the network can learn some features in the situations demonstrated in Fig. \ref{hierarchicalshow}(b) and \ref{hierarchicalshow}(c).
But the situations demonstrated in Fig. \ref{hierarchicalshow}(d) and \ref{hierarchicalshow}(e) may also occur, which are not conducive for the network to learn the features and build a good foundation, and would drag down the learning process.

Therefore, we propose an efficient training strategy.
\editb{The dataset is split into different subsets by difficulty level and fed into the network at different training stages.
At the early training stage, the network is trained with an easy subset (demonstrated in Fig. \ref{hierarchicalshow}(f)).
After the network has learned some feature distribution, the difficulty of training dataset would be increased.
The efficient training strategy allows the network to learn from easy to hard, which we believe is more conducive to the network's learning.}
Specifically, the entire dataset is used for pre-training to obtain an initial model.
The training difficulties of samples are described using the detection loss of this initial model.
The dataset is split into three subsets: easy subset, medium subset and hard subset.
Meanwhile, the efficient training strategy can significantly reduce FLOPs in training process, because the number of samples used in each epoch is less.
The efficient training strategy can also improve the performance and convergence speed of the network.

In addition, one idea for curriculum learning is to reduce the impact of hard samples in the training process, which suggests reducing the probability of hard samples appropriately during training and proposes a strategy to randomly select samples based on their difficulties.
In this paper, we argue that in the early training stage, training the network with only easy samples can build a good foundation for learning strong features.
While in the last training stage, hard samples can improve the detection in subtle areas and further enhance the network's performance.
In the experiments, we compare our efficient training strategy with the random sampling strategy in CD tasks, and the results show the efficient training strategy can achieve better performance.



\subsection{Edge Loss}

\begin{figure}[ht]
\centering
\subfloat[]{
    \includegraphics[width= 0.45\linewidth]{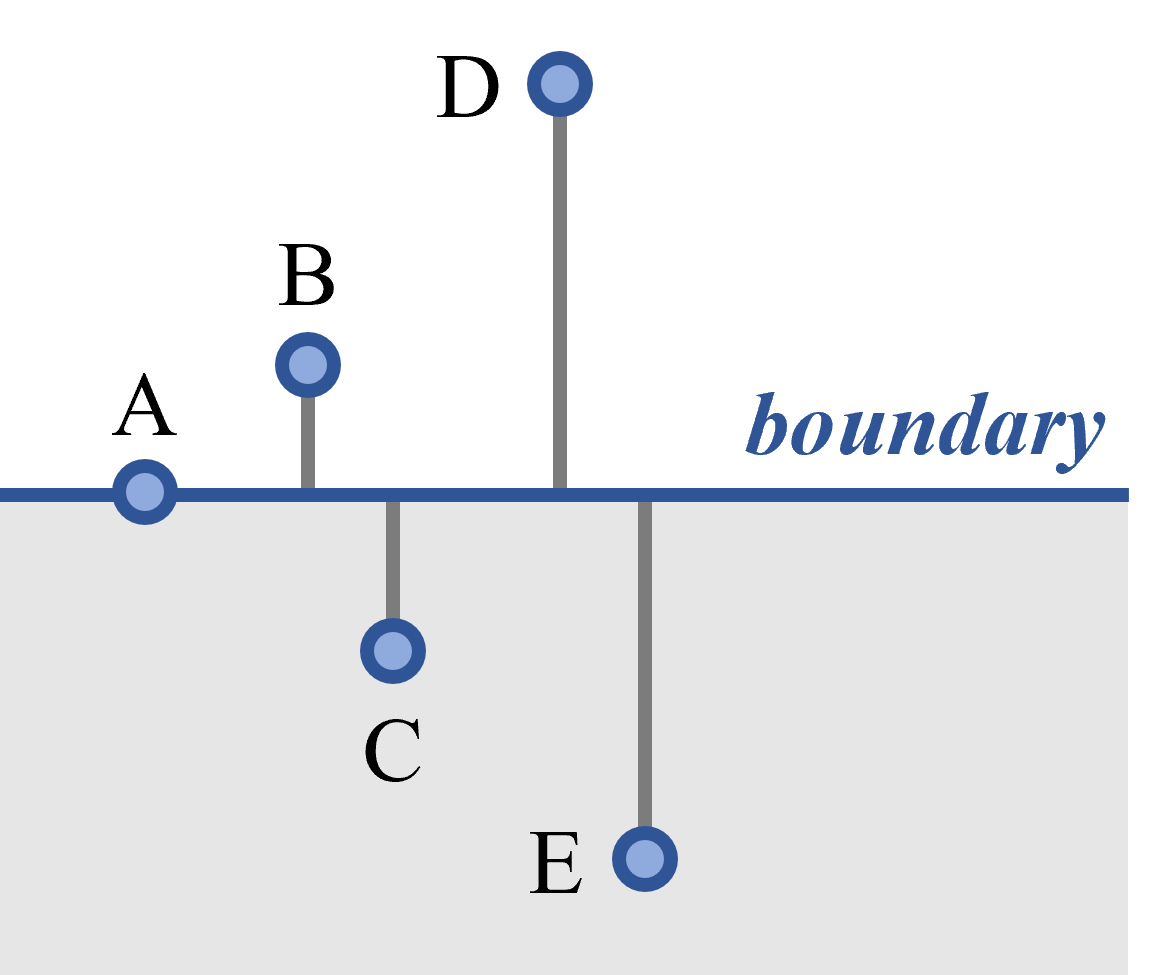}
    \label{boundary1}
}
\subfloat[]{
    \includegraphics[width= 0.35\linewidth]{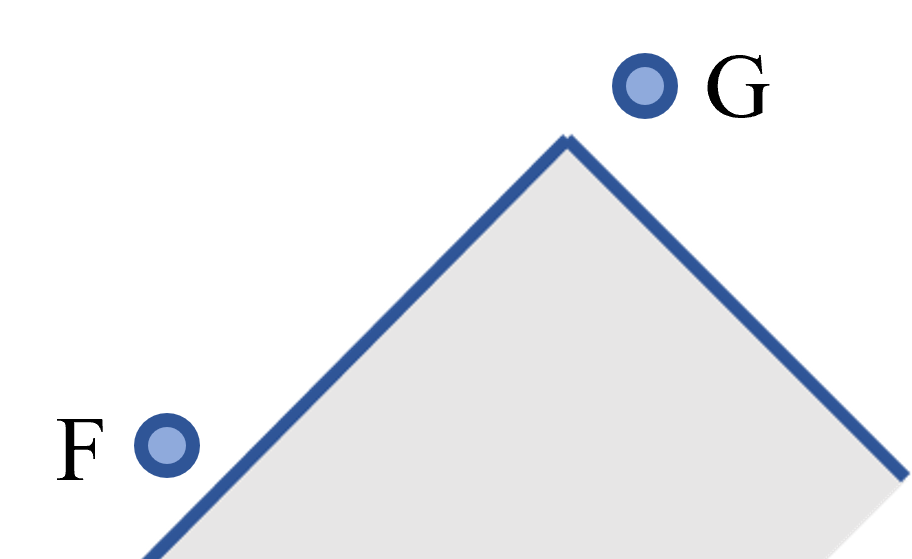}
    \label{boundary2}
}

\subfloat[]{
    \includegraphics[width= 0.35\linewidth]{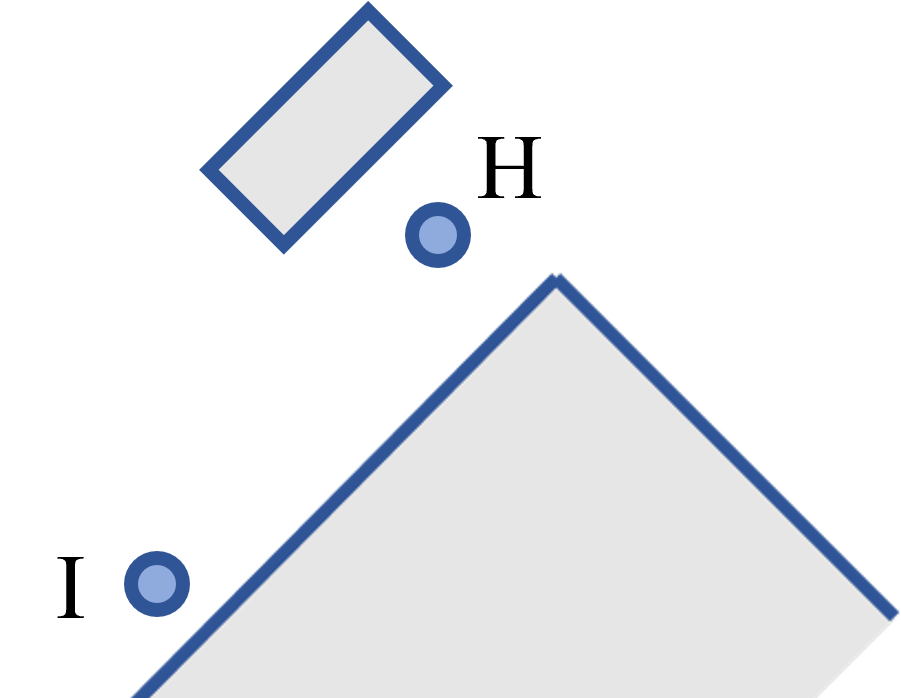}
    \label{boundary3}
}
\subfloat[]{
    \includegraphics[width= 0.35\linewidth]{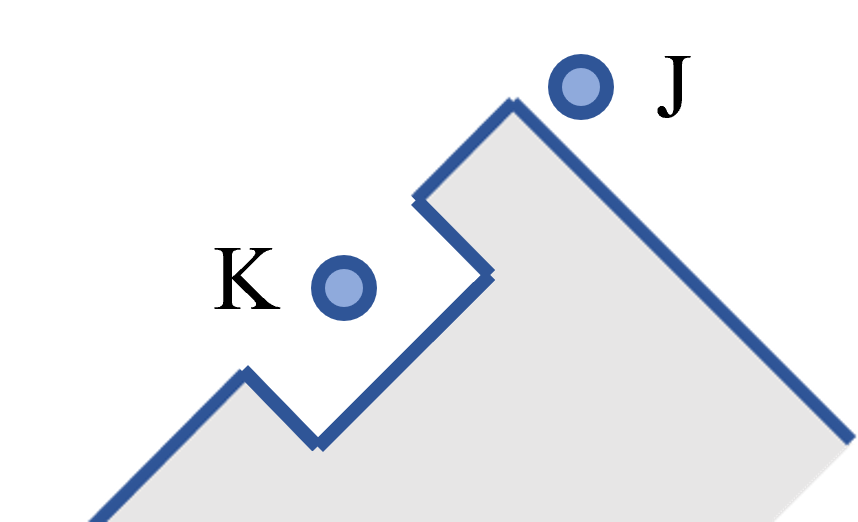}
    \label{boundary4}
}
\caption{(a) shows five cases of points around the straight boundary, where the gray represents the changed area.
(b), (c) and (d) show the other three practical cases.}
\label{boundary}
\end{figure}

Edge loss increases the penalty for errors on the boundary regions by increasing the weights of points located on the boundary during the loss calculation.
It can improve the network's attention to the details in remote sensing CD tasks.
The closer the point is to the boundary, the larger its weight should be, and vice versa.
For a straight boundary, we can roughly divide the points around the boundary into five cases, as shown in Fig. \ref{boundary}(a).
The weight can be determined based on the distance between the point and its boundary.
The weights of each point $w_{\textsf{edge}}({\rm point})$ can be compared as follows: $w_{\textsf{edge}}(A) > w_{\textsf{edge}}(B) \approx w_{\textsf{edge}}(C) > w_{\textsf{edge}}(D) \approx w_{\textsf{edge}}(E)$.
In practice, the scenarios are much more complex.
As shown in Fig. \ref{boundary}(b), in this case, we cannot determine the weight based on its distance from the nearest boundary.
The weight of corner point $G$ should be larger than that of point $F$.
There are many similar cases, as shown in Fig. \ref{boundary}(c) and \ref{boundary}(d).
Obviously, the weight of canyon point $H$ should be larger than point $I$,
and the weight of gap point $K$ should be larger than point $J$.

\begin{figure}[ht]
\centering
\includegraphics[width= 0.4\linewidth]{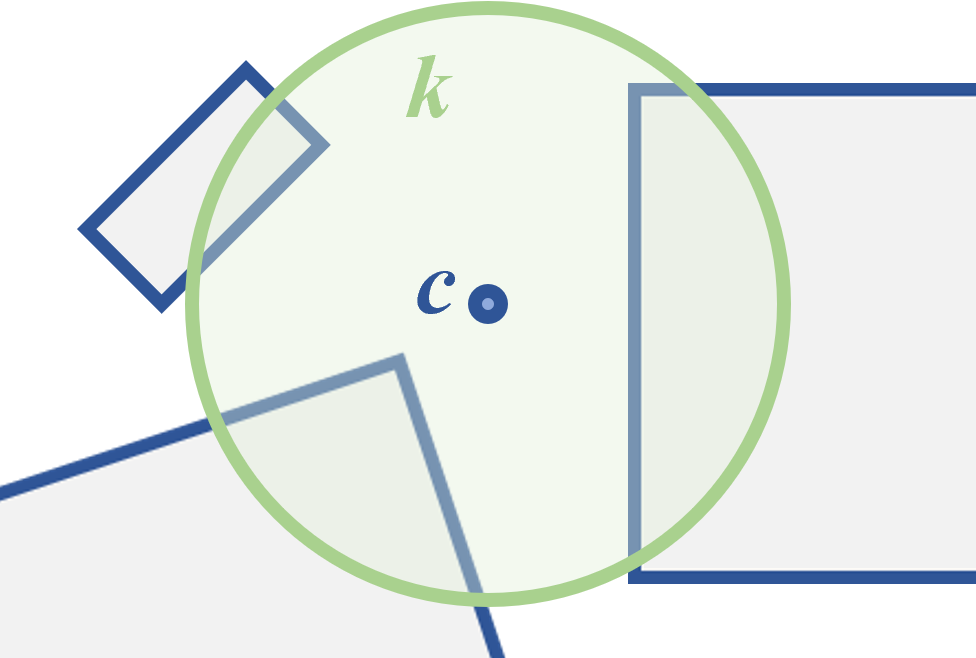}
\caption{The green represents the neighborhood of a point, and the gray represents the changed area.}
\label{boundary5}
\end{figure}

\begin{figure}[ht]
\centering
\subfloat[]{
    \includegraphics[width= 0.3\linewidth]{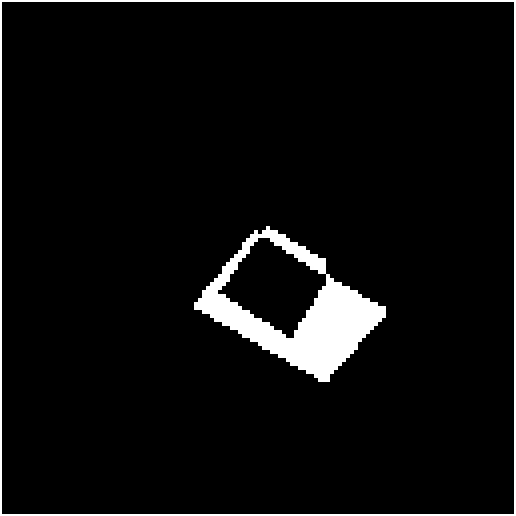}
    \label{edgeloss1}
}
\subfloat[]{
    \includegraphics[width= 0.3\linewidth]{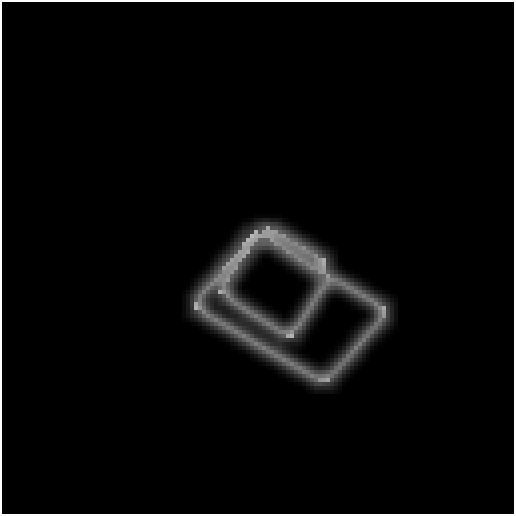}
    \label{edgeloss2}
}
\caption{(a) is the ground truth. 
The white pixels represent the changed area.
(b) is a map of $w_{\textsf{edge}}$.}
\label{edgeloss}
\end{figure}

Therefore, we design a weight $w_{\textsf{edge}}$ describing the changes in the neighborhood of the point, as shown in Fig. \ref{boundary5}.
It can be described as:
\begin{equation}
w_{\textsf{edge}}(c)=\frac{\alpha}{n}\sum_{k \in N} \left| L(k)-L(c)\right|
\end{equation}
where $w_{\textsf{edge}}(c)$ represents the weight of point $c$, $\alpha>0$ is a coefficient, $N$ represents the neighborhood of point $c$ which contains $n$ points, $c$ represents the center of area $N$ and $L[k]$ represents the label value of point $k$.
Since the label value can only be 0 or 1, the above equation is equivalent to:
\begin{equation}
w_{\textsf{edge}}(c)=\alpha \left| \frac{1}{n}\sum_{k \in N}  L(k)-L(c)\right|
\label{edgeeq}
\end{equation}
According to Eq. (\ref{edgeeq}), $w_{\textsf{edge}}$ can be indicated by the difference between the point and the average of its neighborhood.
We combine $w_{\textsf{edge}}$ with the cross entropy loss and define the edge loss as:
\begin{equation}
{\rm EL}(p_t)=- w_{\textsf{edge}} \log (p_t)
\end{equation}
where $p_t$ represents the probability of correct classification.
Fig. \ref{edgeloss} shows an example where $\alpha$ is set to 1.
It can be seen that $w_{\textsf{edge}}$ can quantify the boundary, making the network focus on the details such as boundary regions.
The attention of the network for the boundary regions is likewise increased as $\alpha$ is increased.
The setting of the parameter $\alpha$ will be introduced in the experiment section.


In this paper, we use a hybrid loss function, which is defined as:

\begin{equation}
{\mathcal L}={\mathcal L}_{\rm edge}+{\mathcal L}_{\rm focal}+{\mathcal L}_{\rm dice}
\end{equation}
where ${\mathcal L}_{\rm focal}$ represents Focal loss \cite{lin2017focal} and ${\mathcal L}_{\rm dice}$ represents Dice loss \cite{milletari2016v}.
Edge loss mainly focuses on the details such as boundary regions.
Focal loss and dice loss mainly aim at dealing with class imbalance problems.

\subsection{Network Architecture}

\begin{figure*}[!t]
\centering
\subfloat[Regional Sensing Network (RDP-Net)]{
    \includegraphics[width= 0.7\textwidth]{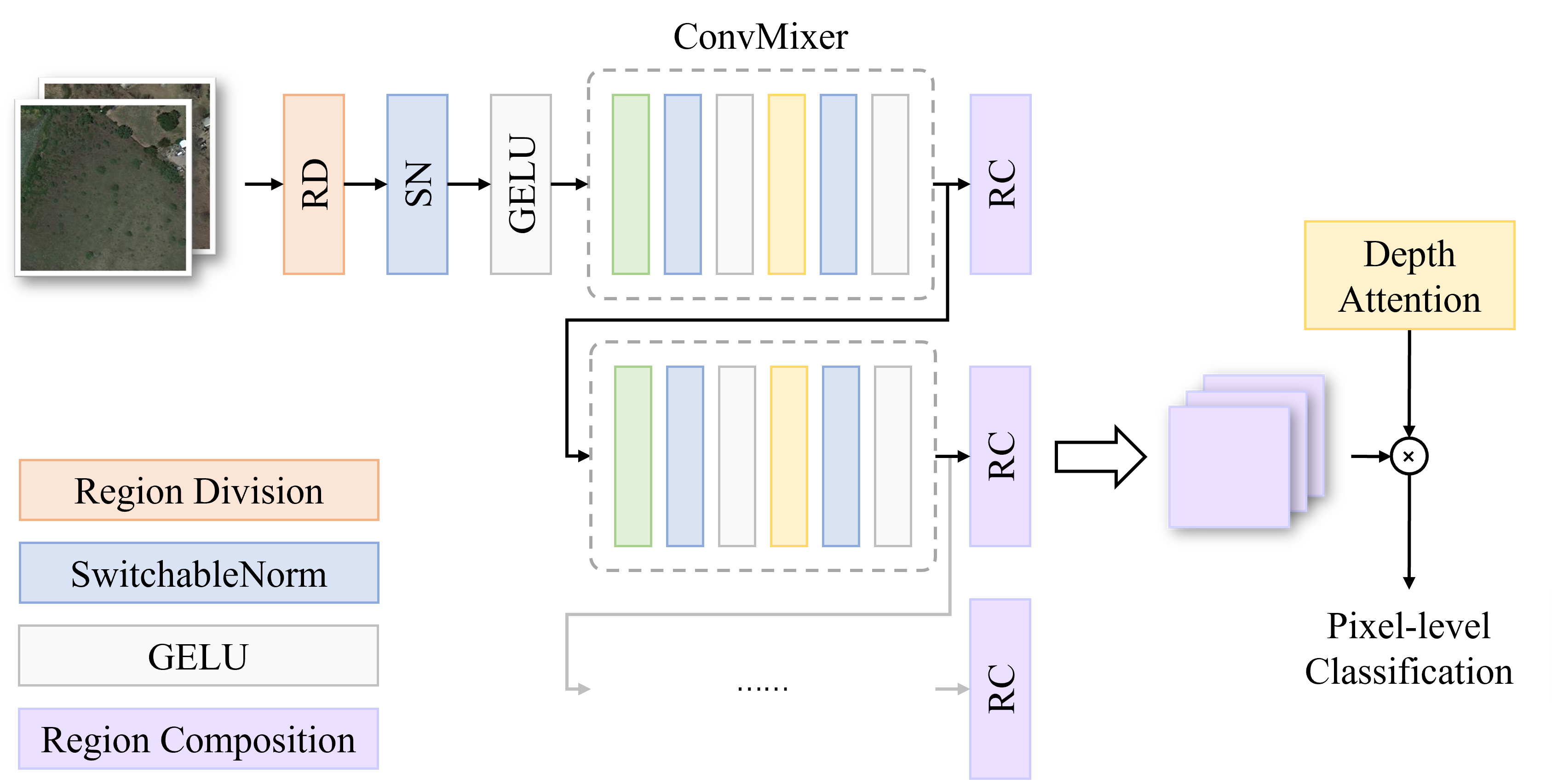}
    \label{RSNet1}
}
\subfloat[ConvMixer]{
    \includegraphics[width= 0.2\textwidth]{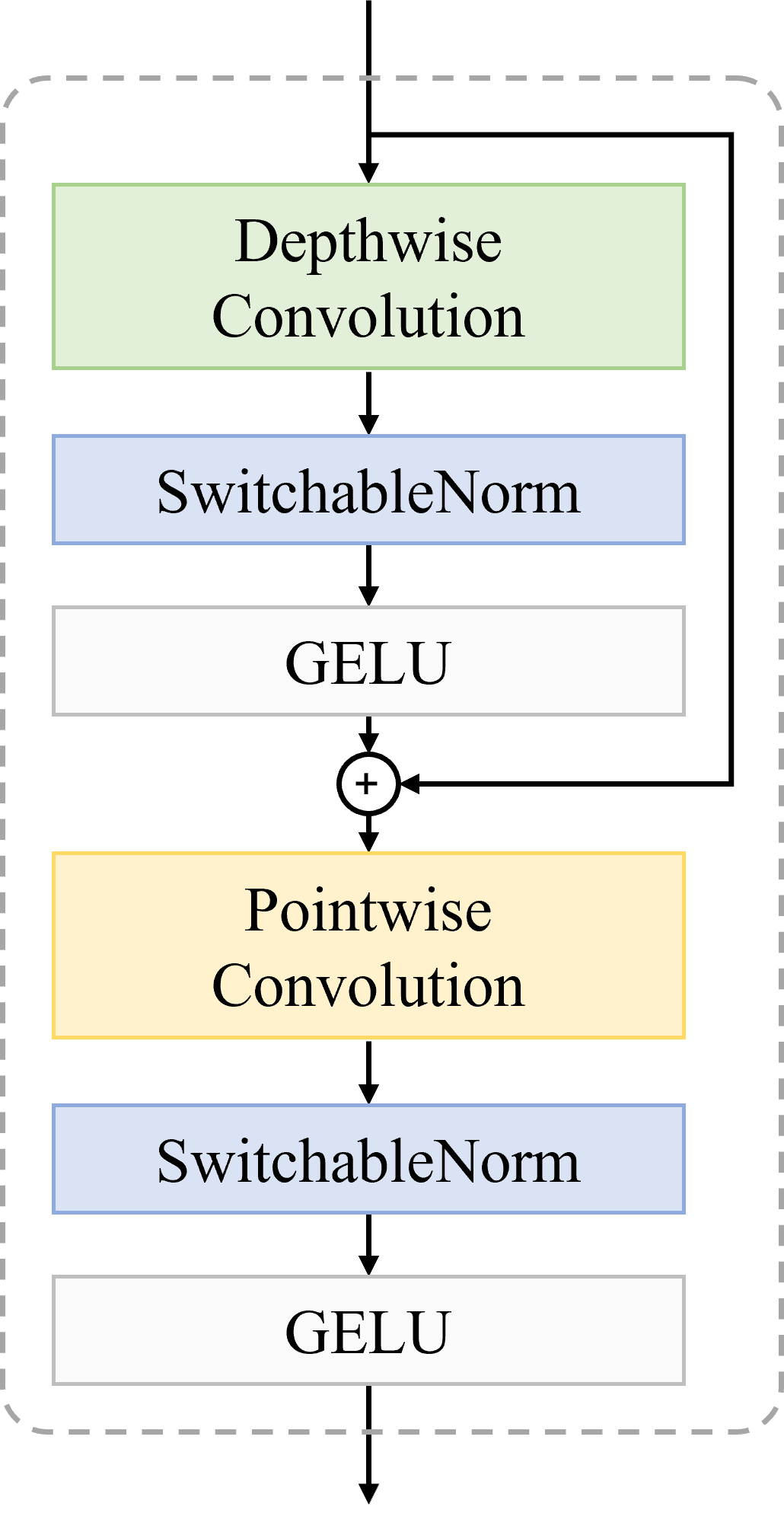}
    \label{RSNet2}
}
\caption{The architecture of the proposed RDP-Net.
(a) is the backbone of RDP-Net.
(b) is ConvMixer.}
\label{RSNet}
\end{figure*}

Our model, based on a brand new backbone, named RDP-Net, mainly consists of four components, including a region division layer, a fully-convolutional block named ConvMixer, and a region composition layer, as shown in Fig. \ref{RSNet}(a).
For change detection tasks, detailed local information is more valuable than global context, especially for the change detection of boundary regions and small areas.
So, we design a region division layer to slice the input image into several small patches by region.
ConvMixer allows us to explore the local information in the patch while perceiving the global context.
The region composition layer will compose the patches. 
Finally, a tiny depth attention module is added to fuse the multiple depth outputs and obtain the change detection result.

Specifically, the region division layer is used to slice the input image into several patches according to the region.
In order to maintain the amount of data without loss and preserve all the detailed local information, the region division layer with region size ($p \times p$) and input image size ($c \times h \times w$) can be implemented as convolution with $c_{in}$ input channels, $c_{out}$ ($=c_{in} \times(h/p\times w/p)$) output channels, kernel size $p$, and stride $p$:
\begin{equation*}
\sigma \left\{ {\rm SN}( {\rm Conv} \left( c_{in}, c_{out}, {\rm stride}=p, {\rm kernel}=p \right) ) \right\}
\end{equation*}
where $c, h, w$ represent the channels, height and width of the input image, SN() represents Switchable Normalization \cite{luo2018differentiable}, $\sigma \{\}$ represents Gaussian Error Linear Units \cite{hendrycks2016gaussian}.
Switchable Normalization allows the network to choose a better normalization method for each layer, so that we can pay attention to the module design of the network.
GELU combines the idea of dropout \cite{hinton2012improving} and ReLU \cite{nair2010rectified}, making the network training more robust.

The ConvMixer block \cite{anonymous2022patches} is used as the backbone of our network.
As shown in Fig. \ref{RSNet}(b), it consists of depthwise convolution followed by pointwise convolution.
The depthwise convolution is used to explore information between different patches.
The pointwise convolution is used to explore the information of each patch.
The residual connection ensures that it still mainly focuses on the details inside a path, when the network explores information between patches, which is important for detail-sensitive change detection tasks.
In this paper, we introduce ConvMixer into the change detection task.
Existing methods gradually lose information as the network deepens.
The ConvMixer makes the resolution of the network remain the same and retains detailed local information, so it is suitable for change detection tasks that require more attention to detail.

The region composition layer is used to compose patches together according to the region and obtain the pixel-level classification feature map.
The region composition layer can be implemented as convolution-transpose with $c_{out}$ ($=c_{in} \times(h/p\times w/p)$) input channels, $out\_ch$ output channels, kernel size $p$, and stride $p$:
\begin{equation*}
\sigma \left\{ {\rm SN}( {\rm ConvTrans} \left( c_{out}, out\_ch, {\rm stride}=p, {\rm kernel}=p \right) ) \right\}
\end{equation*}

The tiny depth attention module is used to suppress semantic gaps and localization differences.
It is a 1d learnable weight vector with length $out\_ch \times depth$, where $depth$ represents the depth of the backbone.

RDP-Net source code is released at \url{https://github.com/Chnja/RDPNet}.

\section{Experiment and Result Analysis}

\subsection{Dataset}

The experiment was conducted on two datasets named CDD \cite{lebedev2018change} and LEVIR-CD \cite{Chen2020}, two of the most common datasets in remote sensing change detection.

\textit{CDD} dataset consists of seven image pairs of $4725 \times 2200$ pixels and four image pairs of $1900 \times 1000$ pixels.
The spatial resolution ranges from 3 to 100 cm per pixel, and the seasons vary widely.
We cut each image pair into $256 \times 256$ pixel patches without overlapping and ultimately obtained 10000 training sets and 3000 validation sets.

\textit{LEVIR-CD} dataset consists of 637 image pairs of $1024 \times 1024$ pixels.
We cut each image pair into $256 \times 256$ pixel patches without overlapping and ultimately obtained 3167 training sets and 436 validation sets.

\subsection{Implementation Details}

We implement RDP-Net using Pytorch framework.
The depth of ConvMixer is set to 6, $out\_ch$ is set to 32, and the size of the neighborhood in edge loss is set to 7.
The training dataset is divided into three subsets: easy subset, medium subset and hard subset.
The ratio of the three subsets is set to 4:2:3 empirically.
Different subsets are fed to the network at the 30th, 60th and 90th epochs.
The learning rate is set to 1e-3 and decays by 0.8 every 15 epochs.
In the training process, the batch size is set to 16, and AdamW \cite{loshchilov2017decoupled} is applied as an optimizer.
The parameter $\alpha$ in edge loss is set to 1.
We conduct experiments on a single NVIDIA RTX3090 and train for 200 epochs.

And in the experiments, we use three indicators for the evaluation of quantitative metrics: Precision, Recall, and F1-Score.

\subsection{Comparison With State-of-the-Art Methods}

We compare our method with FC-EF, FC-Siam-conc, FC-Siam-diff \cite{daudt2018fully}, UNet++\_MSOF \cite{peng2019end}, DASNet \cite{chen2020dasnet}, STANet \cite{chen2020spatial}, SNUNet-CD \cite{fang2021snunet}, AGCDetNet \cite{song2021agcdetnet} and SiamixFormer\cite{ghaderi2022siamixformer}.
They are representative of deep learning-based methods in the field of change detection.
\editb{FC-EF, FC-Siam-conc and FC-Siam-diff \cite{daudt2018fully} are the baseline model for change detection, they are the promotion of U-Net \cite{ronneberger2015u}.}
Unet++\_MSOF uses the multiple side outputs fusion (MSOF) from U-Net++ \cite{zhou2018unet++} for deep supervision.
DASNet uses a contrastive method with a dual attention mechanism.
\editb{STANet and AGCDetNet can provide a multi-scale understanding of the difference map.} 
SNUNet-CD uses the ensemble channel attention module to fuse the multiple side outputs of U-Net++, and its inputs come from a Siamese network.
\editb{SiamixFormer uses the Transformer with a Siamese network.}

\begin{table*}[ht]
\caption{Results on CDD dataset and LEVIR-CD dataset}
\label{cddsota}
\centering
\begin{tabular}{ccccccccc}
\hline
\hline
\multirow{2}*{Method / Channel} & \multirow{2}*{Params(M)} & \multirow{2}*{\editc{FLOPs(G)}} & \multicolumn{3}{c}{CDD dataset} & \multicolumn{3}{c}{LEVIR-CD dataset}\\
\cline{4-9}
 & & & Precision & Recall & F1 & Precision & Recall & F1\\
\hline
FC-EF & 1.83 & \editc{2.1} & 0.598 & 0.708 & 0.648 & 0.776 & 0.712 & 0.735\\
FC-Siam-conc & 2.03 & \editc{5.3} & 0.679 & 0.723 & 0.700 & 0.899 & 0.775 & 0.828\\
FC-Siam-diff & 1.83 & \editc{4.7} & 0.715 & 0.685 & 0.699 & 0.883 & 0.785 & 0.826\\
UNet++\_MSOF / 16 & 2.75 & \editc{11.4} & 0.924 & 0.879 & 0.901 & 0.907 & 0.880 & 0.893\\
UNet++\_MSOF / 32 & 11.00 & \editc{45.4} & 0.946 & 0.939 & 0.943 & 0.906 & 0.886 & 0.894\\
DASNet & 48.22 & \editc{100.7} & 0.914 & 0.925 & 0.919 & 0.811 & 0.788 & 0.799\\
\editb{STANet} & 16.93 & - & 0.832 & 0.928 & 0.877 & 0.831 & 0.887 & 0.834\\
SNUNet-CD / 16 & 3.31 & \editc{13.8} & 0.938 & 0.935 & 0.936 & \textbf{0.917} & 0.871 & 0.892\\
SNUNet-CD / 32 & 13.21 & \editc{54.8} & 0.961 & 0.965 & 0.963 & 0.907 & 0.887 & 0.896\\
\editb{AGCDetNet} & 43.05 & \editc{79.6} & 0.960 & 0.971 & 0.965 & 0.891 & 0.888 & 0.899\\
\editb{SiamixFormer-0} & 6.80 & - & - & - & 0.921 & - & - & 0.895\\
\editb{SiamixFormer-1} & 26.20 & - & - & - & 0.928 & - & - & 0.893\\
\textbf{RDP-Net} & \textbf{1.70} & \editc{\textbf{27.2}} & \textbf{0.967} & \textbf{0.977} & \textbf{0.972} & 0.915 & \textbf{0.888} & \textbf{0.901}\\
\hline
\hline
\end{tabular}
\end{table*}


Table \ref{cddsota} report the comparisons of detection accuracy and the number of parameters.
Our proposed RDP-Net can get better performance than other SOTA change detection methods, and it only has 1.70 M parameters.
CDD dataset contains some detailed changes. Although the overall task is not difficult, it is still a challenge to further improve accuracy.
It can be seen that our network has a significant improvement.
Our RDP-Net only uses 1.7M parameters ($\approx 12.9\% \times {\rm 13.2M}$) to achieve a better result (F1 = 0.972), which undoubtedly has advantages.
Compared with SNUNet-CD / 16 with twice parameters, the F1 is increased by 3.6\%.
LEVIR-CD dataset is mainly for building change detection tasks.
The label in LEVIR-CD dataset mainly refers to the general location and shape of the building, which is not as fine as the CDD dataset.
Our RDP-Net does not perform well on this dataset, but still achieves better performance with fewer parameters.
We believe that on datasets with more accurate labels, our method will have a more obvious improvement effect.

\begin{figure*}[!t]
\centering
\subfloat[]{
    \includegraphics[width= 0.13\textwidth]{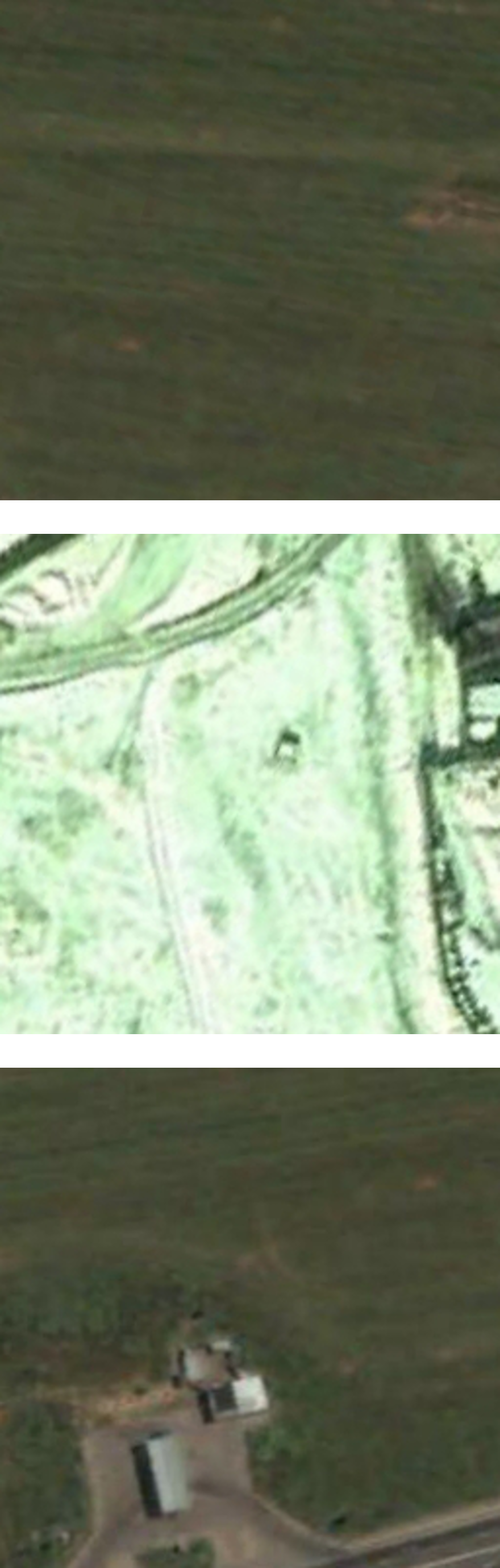}
}
\subfloat[]{
    \includegraphics[width= 0.13\textwidth]{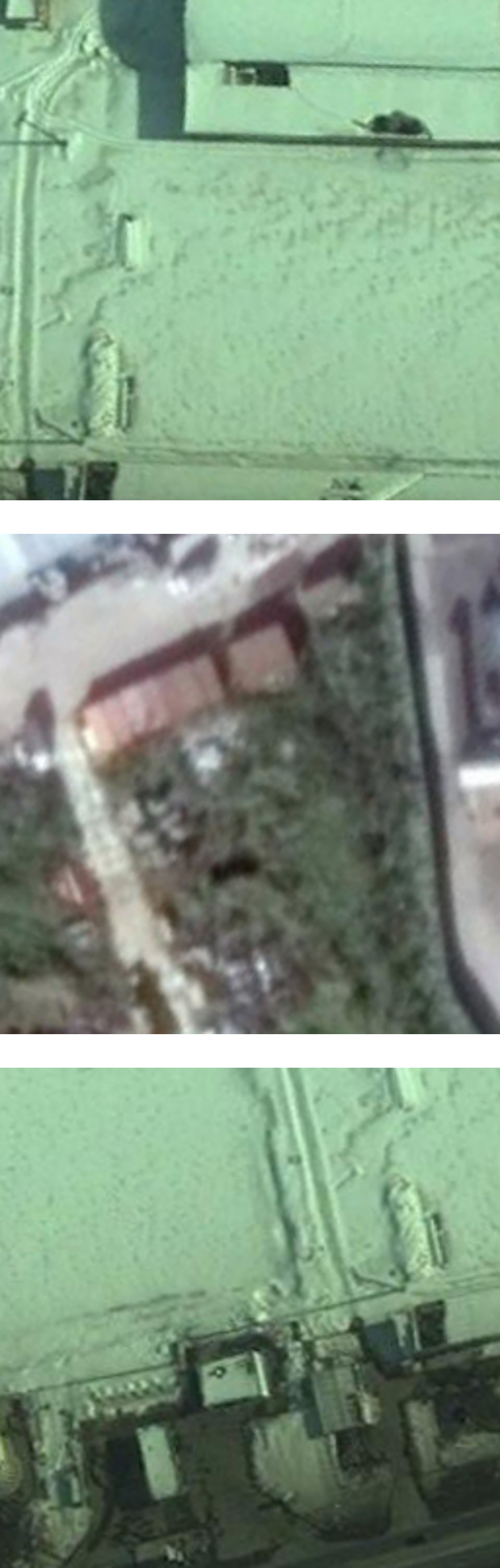}
}
\subfloat[]{
    \includegraphics[width= 0.13\textwidth]{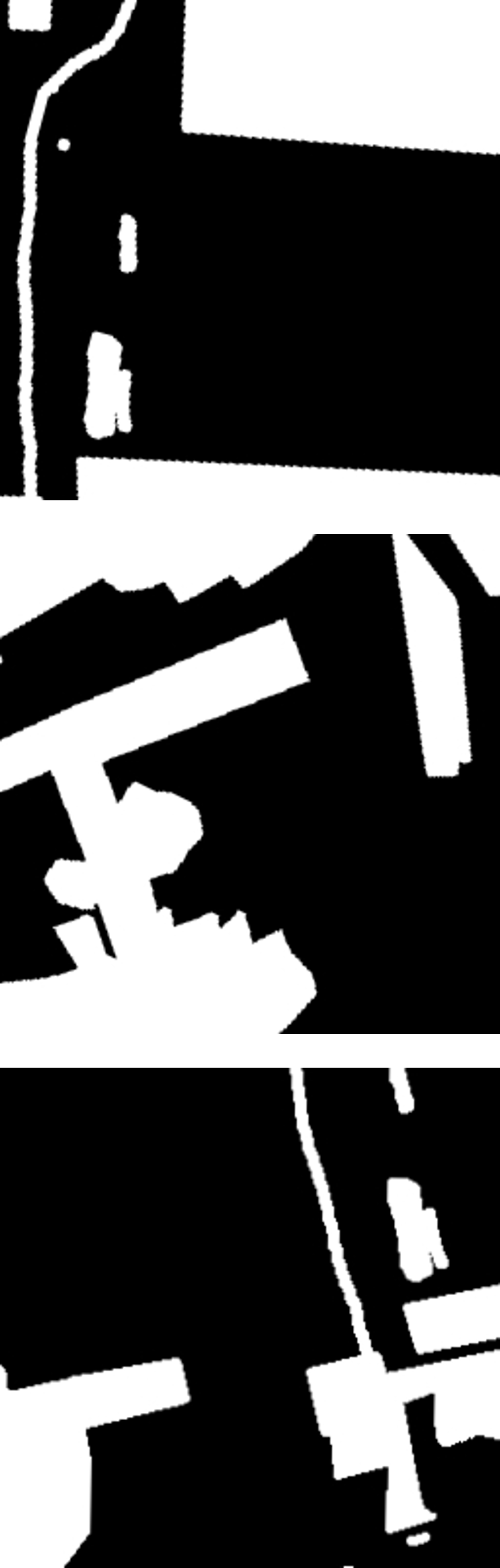}
}
\subfloat[]{
    \includegraphics[width= 0.13\textwidth]{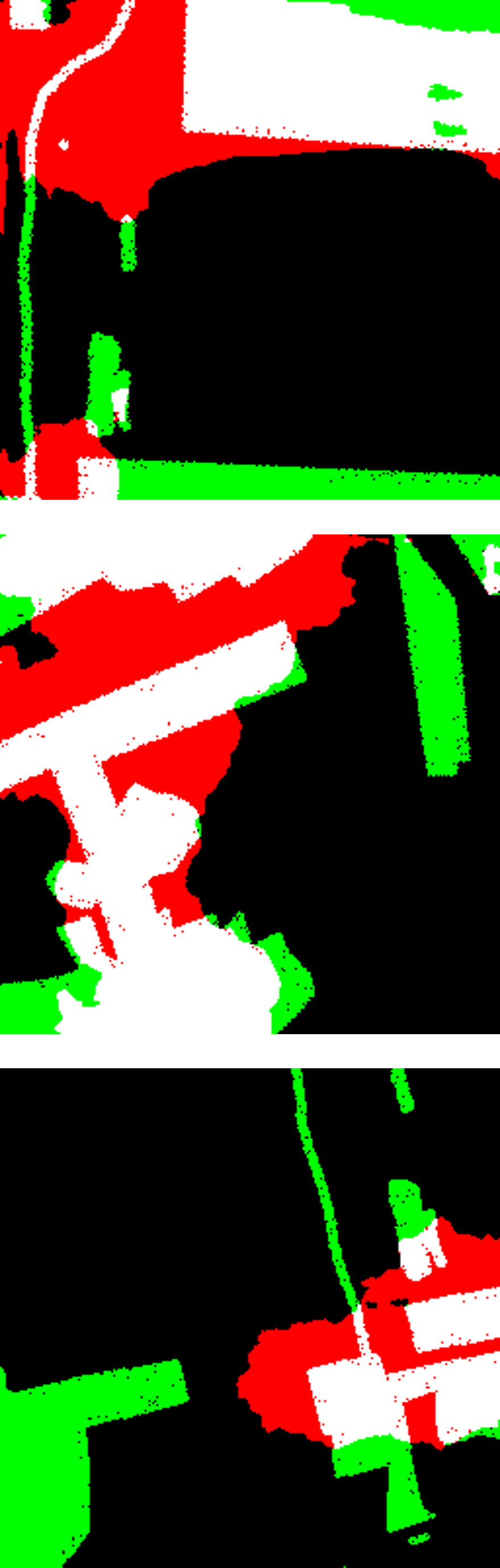}
}
\subfloat[]{
    \includegraphics[width= 0.13\textwidth]{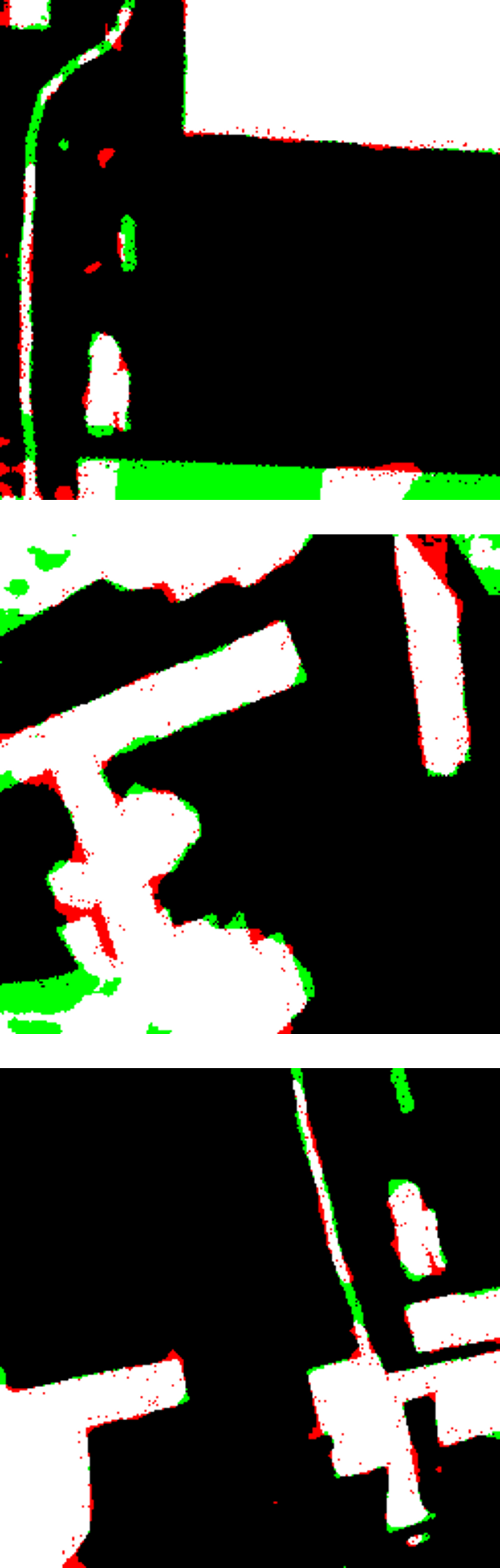}
}
\subfloat[]{
    \includegraphics[width= 0.13\textwidth]{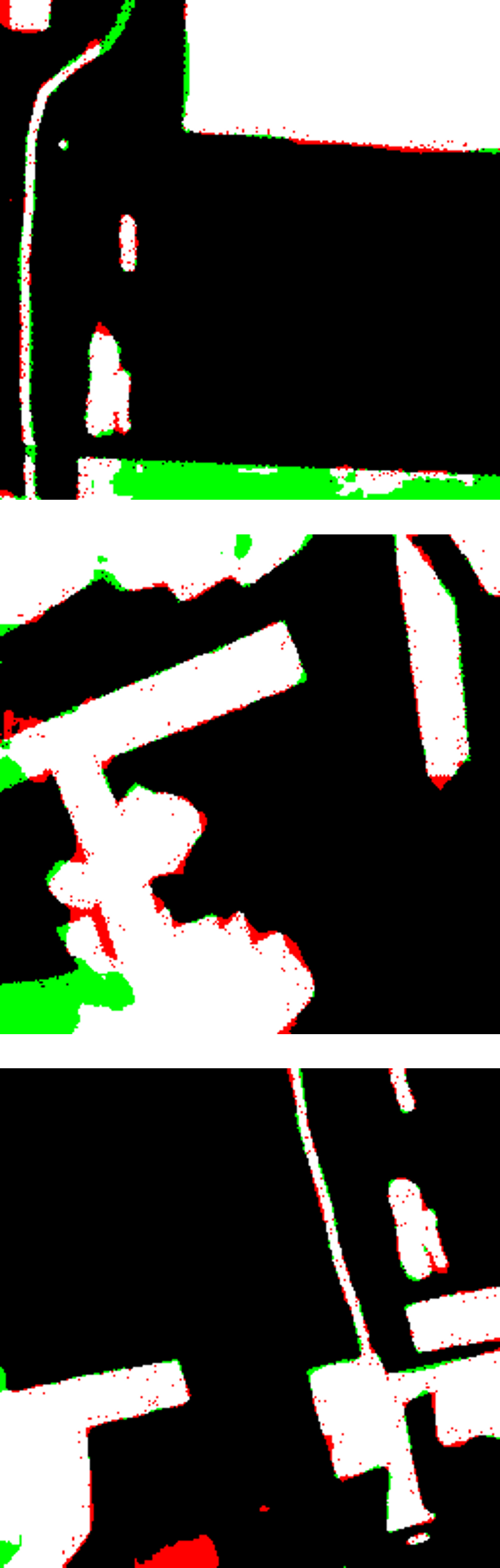}
}
\subfloat[]{
    \includegraphics[width= 0.13\textwidth]{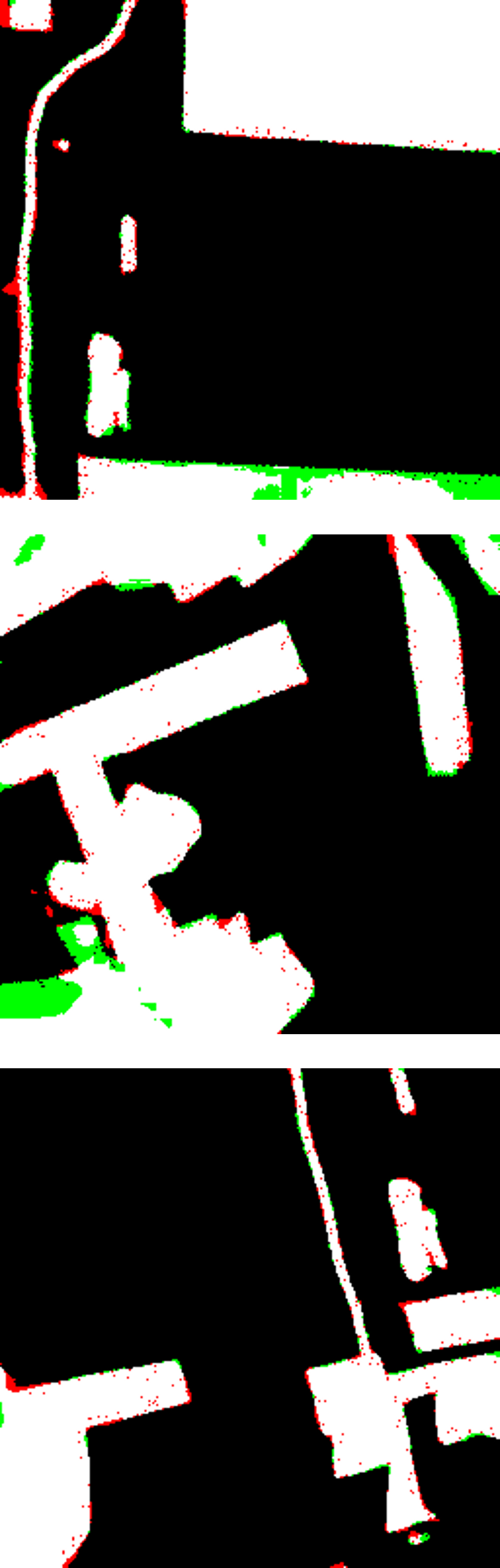}
}

\subfloat[]{
    \includegraphics[width= 0.13\textwidth]{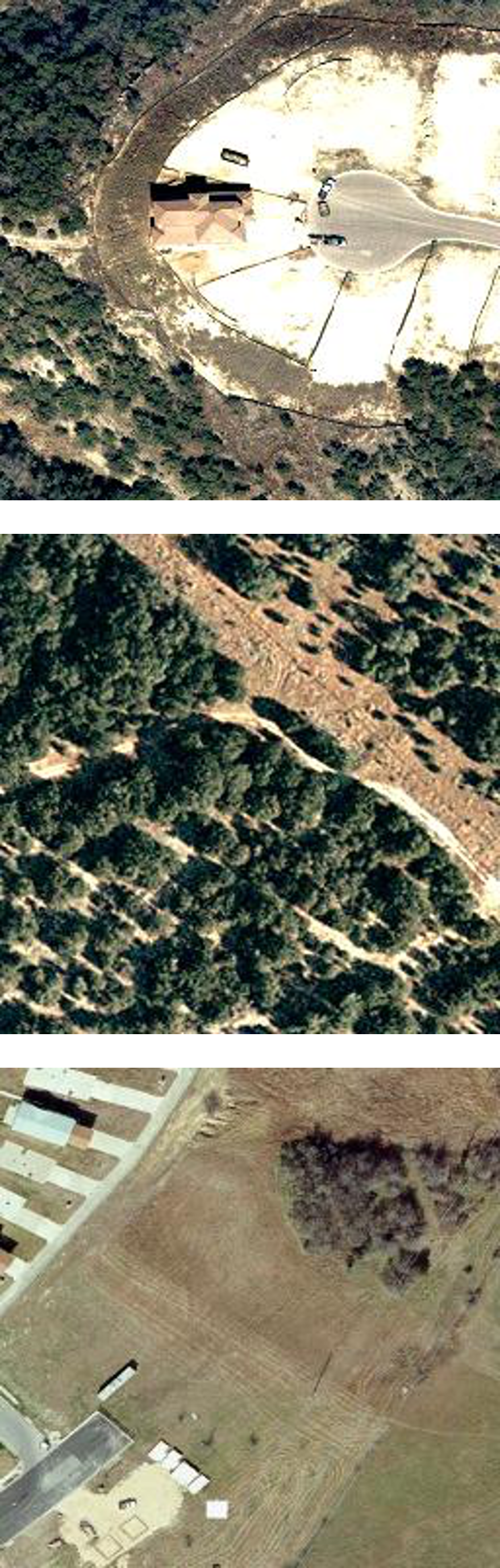}
}
\subfloat[]{
    \includegraphics[width= 0.13\textwidth]{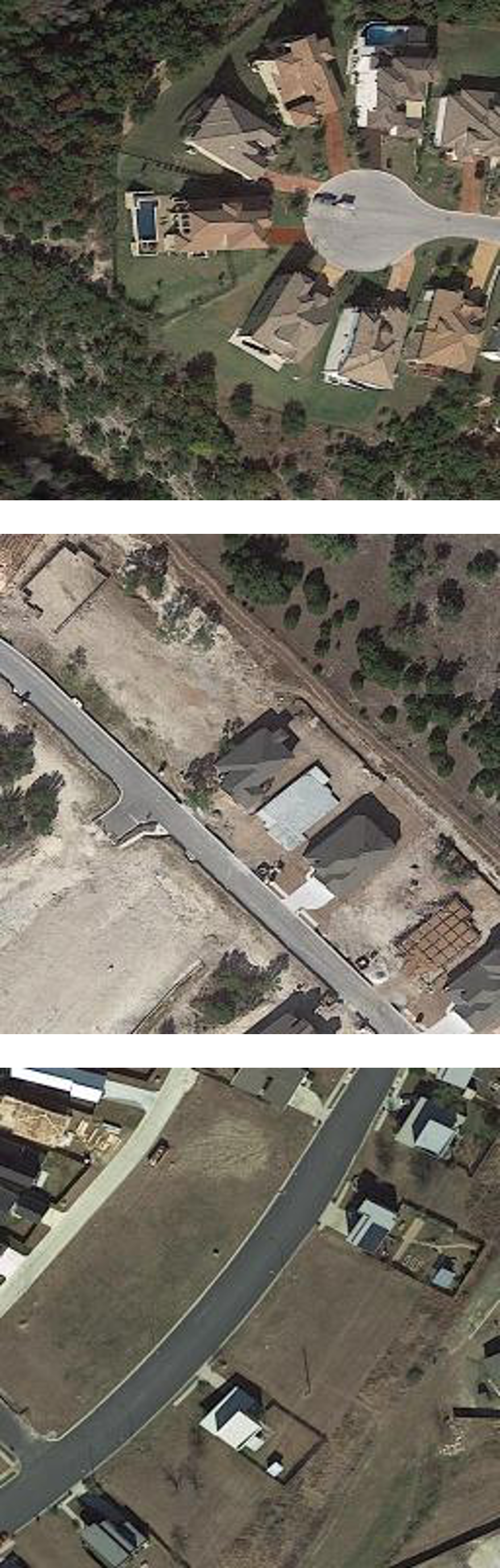}
}
\subfloat[]{
    \includegraphics[width= 0.13\textwidth]{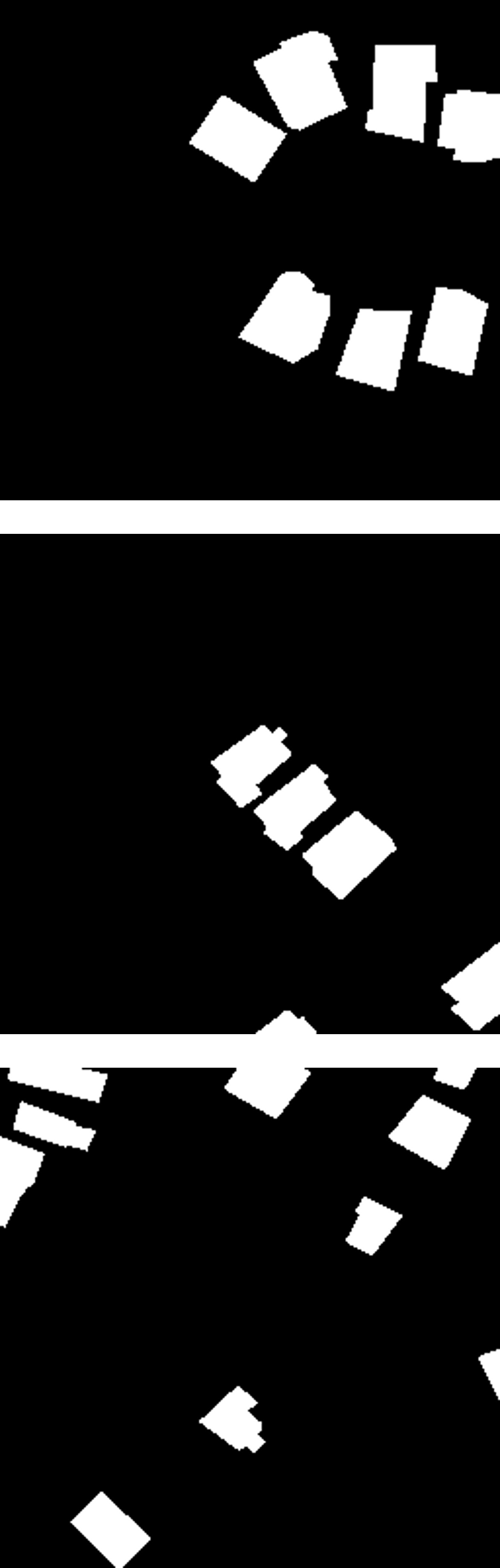}
}
\subfloat[]{
    \includegraphics[width= 0.13\textwidth]{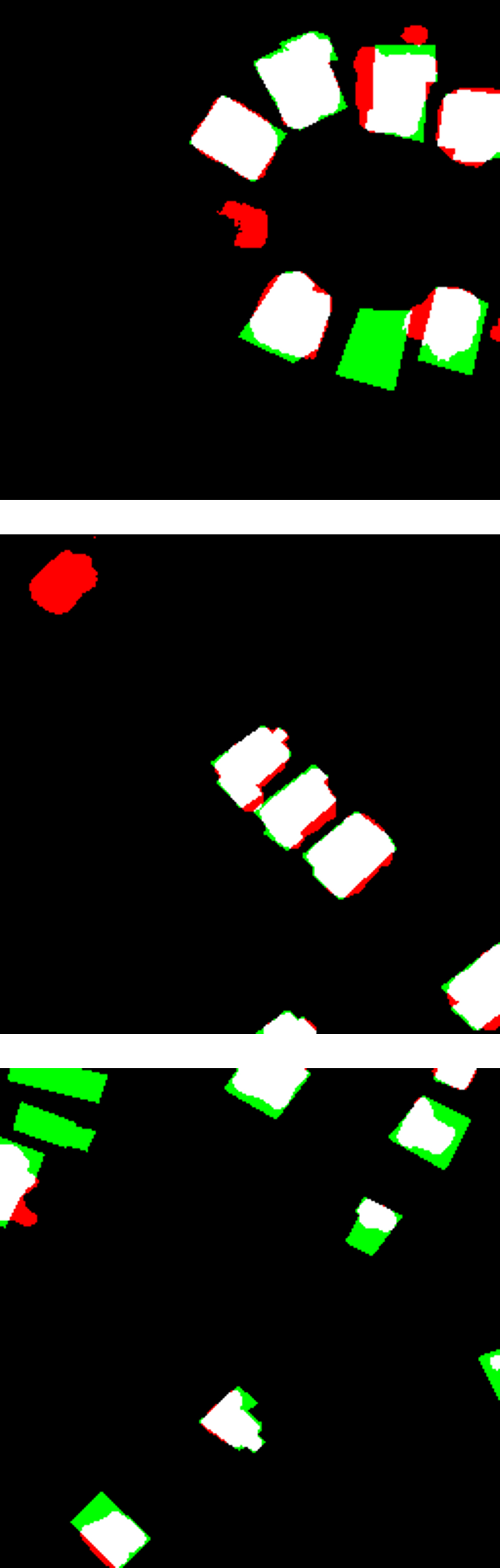}
}
\subfloat[]{
    \includegraphics[width= 0.13\textwidth]{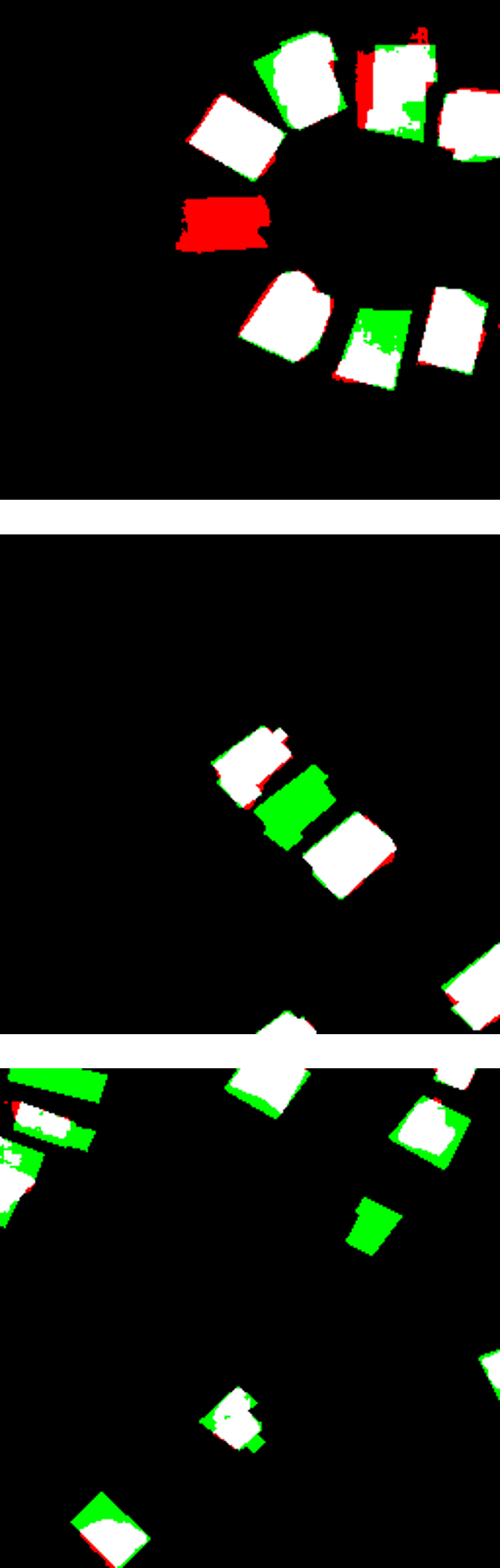}
}
\subfloat[]{
    \includegraphics[width= 0.13\textwidth]{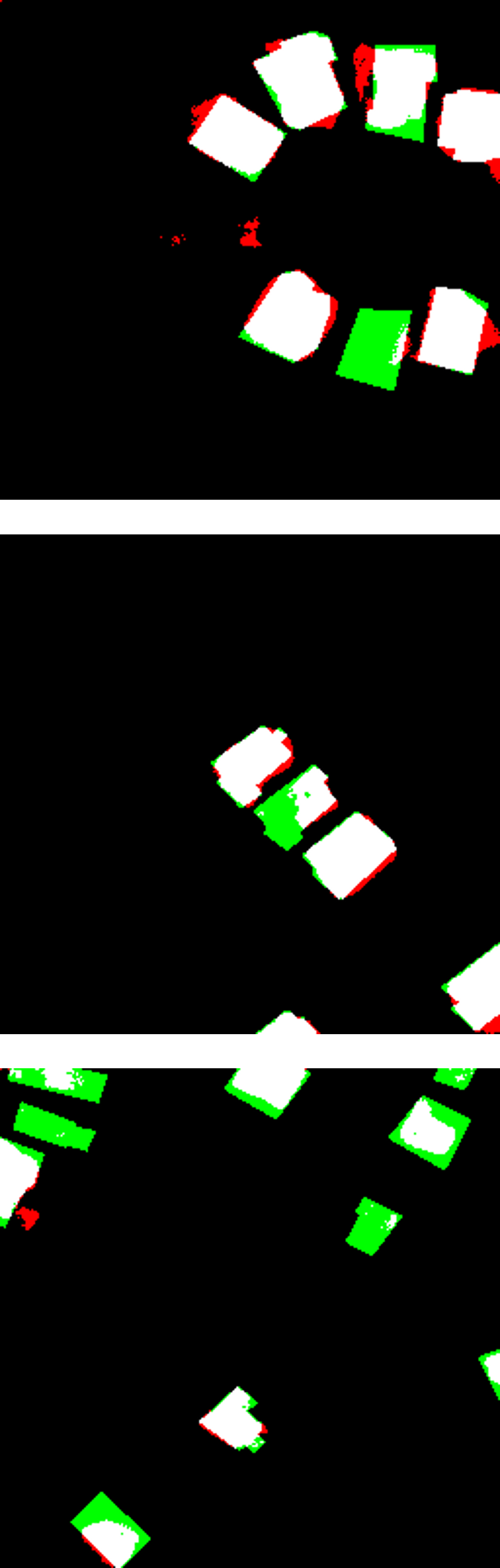}
}
\subfloat[]{
    \includegraphics[width= 0.13\textwidth]{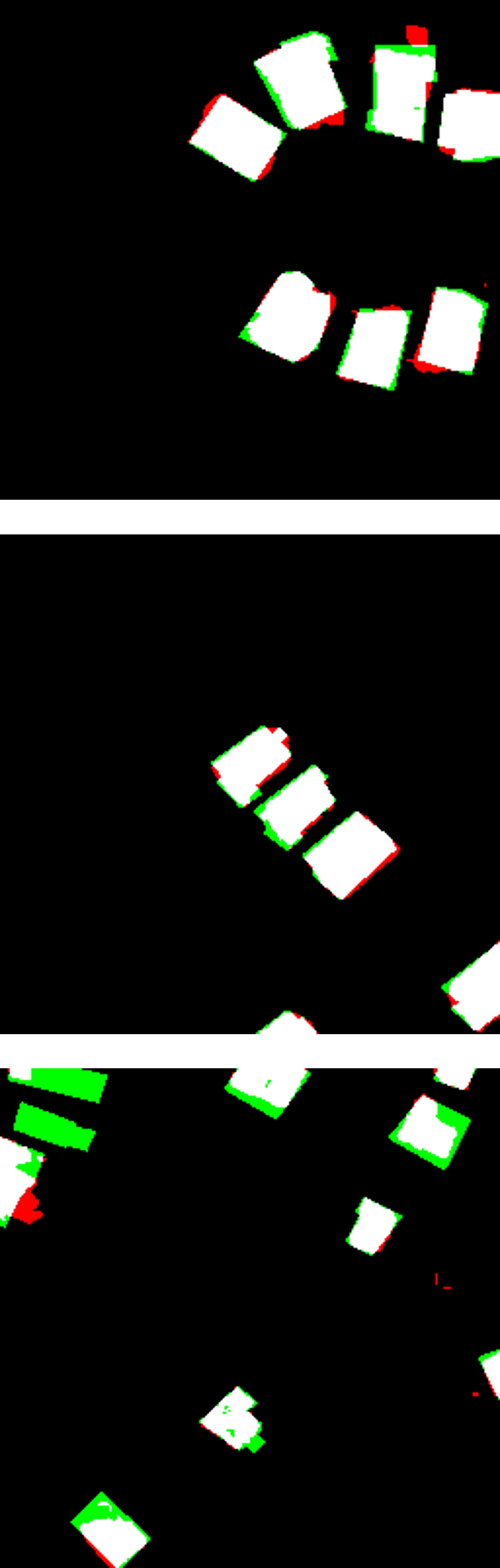}
}
\caption{(a)-(g) are results on CDD dataset.
(h)-(n) are results on LEVIR-CD dataset.
(a), (b), (h) and (i) are the original images.
(c) and (j) are the ground truth.
The results of (d) (k) FC-Siam-diff,
(e) (l) UNet++\_MSOF with 32 channels,
(f) (m) SNUNet-CD with 32 channels,
(g) (n) our RDP-Net.
The false positives and false negatives are indicated by red and green, respectively.
Other colors represent true positives.}
\label{final}
\end{figure*}

In addition, Fig. \ref{final} shows some detection results from the validation set of the CDD and LEVIR-CD datasets.
The false positives and false negatives are indicated by red and green, respectively.
Other colors represent true positives.

\subsection{Ablation Study}

To evaluate the RDP-Net, edge loss and efficient training strategy, a couple of ablation experiments were conducted.
Table \ref{Ablation} and \ref{Ablation2} report the detection accuracy.

\begin{table}[ht]
\caption{Ablation Experiment on CDD Dataset}
\label{Ablation}
\centering
\begin{tabular}{cccc}
\hline
\hline
Method & Precision & Recall & F1\\
\hline
w/o Edge loss, efficient training & 0.966 & 0.965 & 0.965\\
w/o efficient training & 0.965 & 0.970 & 0.967\\
RDP-Net & 0.967 & 0.977 & 0.972\\
\hline
\hline
\end{tabular}
\end{table}

\begin{table}[ht]
\caption{Ablation Experiment on LEVIR-CD Dataset}
\label{Ablation2}
\centering
\begin{tabular}{cccc}
\hline
\hline
Method & Precision & Recall & F1\\
\hline
w/o Edge loss, efficient training & 0.906 & 0.872 & 0.887\\
w/o efficient training & 0.905 & 0.882 & 0.892\\
RDP-Net & 0.915 & 0.888 & 0.901\\
\hline
\hline
\end{tabular}
\end{table}


\begin{figure*}[ht]
\centering
\subfloat[]{
    \includegraphics[width= 0.13\linewidth]{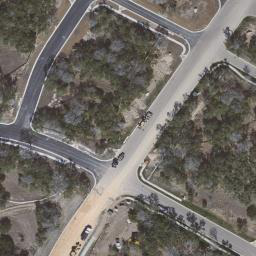}
}
\subfloat[]{
    \includegraphics[width= 0.13\linewidth]{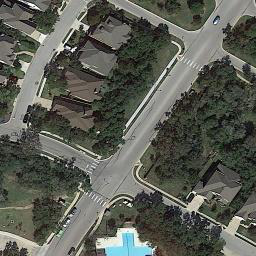}
}
\subfloat[]{
    \includegraphics[width= 0.13\linewidth]{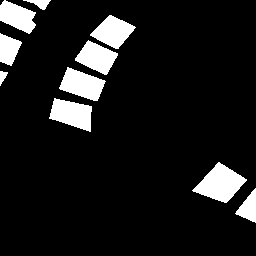}
}
\subfloat[]{
    \includegraphics[width= 0.13\linewidth]{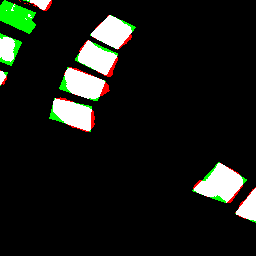}
}
\subfloat[]{
    \includegraphics[width= 0.13\linewidth]{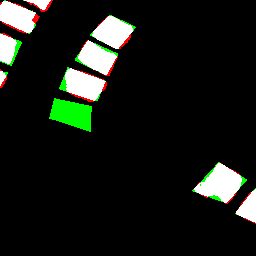}
}
\subfloat[]{
    \includegraphics[width= 0.13\linewidth]{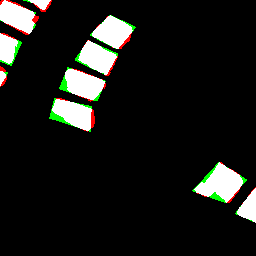}
}
\subfloat[]{
    \includegraphics[width= 0.13\linewidth]{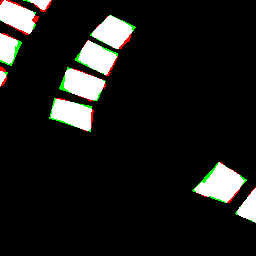}
}

\subfloat[]{
    \includegraphics[width= 0.13\linewidth]{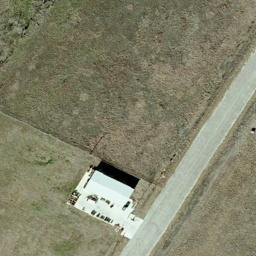}
}
\subfloat[]{
    \includegraphics[width= 0.13\linewidth]{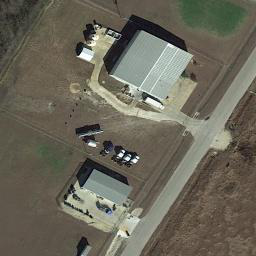}
}
\subfloat[]{
    \includegraphics[width= 0.13\linewidth]{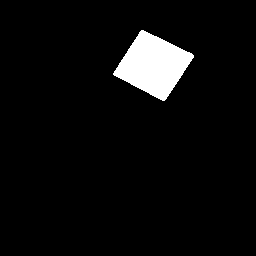}
}
\subfloat[]{
    \includegraphics[width= 0.13\linewidth]{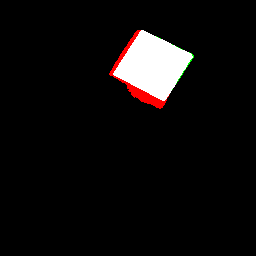}
}
\subfloat[]{
    \includegraphics[width= 0.13\linewidth]{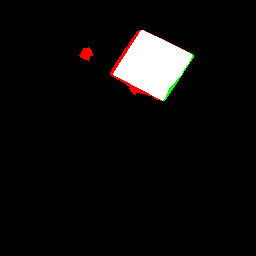}
}
\subfloat[]{
    \includegraphics[width= 0.13\linewidth]{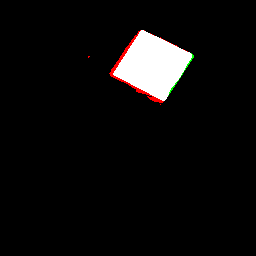}
}
\subfloat[]{
    \includegraphics[width= 0.13\linewidth]{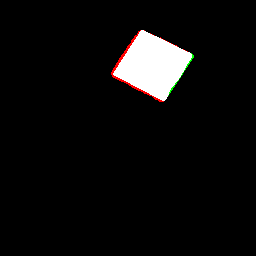}
}
\caption{(a), (b), (h) and (i) are the original input images.
(c) and (j) are the ground truth.
The results of (d) (k) SNUNet-CD with 32 channels,
(e) (l) our RDP-Net without edge loss and efficient training strategy,
(f) (m) our RDP-Net without efficient training strategy.
(g) (n) our RDP-Net.
The false positives and false negatives are indicated by red and green, respectively.
Other colors represent true positives.}
\label{ablationimg}
\end{figure*}

The experimental results show that the addition of the Edge loss is beneficial to improving the detection accuracy by 0.5\% recall and 0.2\% F1 on CDD dataset; 1\% recall and 0.5\% F1 on LEVIR-CD dataset.
Fig. \ref{ablationimg} shows the contribution of edge loss, that the model can achieve a better performance in boundary regions with edge loss.
The efficient training strategy allows the network to learn from easy to hard. When we use it to train the network, the detection accuracy is improved by 0.7\% recall and 0.5\% F1 on CDD dataset; 1\% precision, 0.6\% recall and 0.9\% F1 on LEVIR-CD dataset.

We can infer from the experiment that our improvements can increase recall without reducing precision, which enables the network to detect as many change areas as possible without increasing the false detection rate.
Meanwhile, the experiment can also prove that the architecture of RDP-Net is very efficient, which enables us to achieve a performance close to SOTA even without edge loss and the efficient training strategy.

\subsection{Experiment in Edge Loss}

\editc{To test the effect of parameter $\alpha$ in edge loss, we selected five values from 0.1 to 5, and tested them on CDD dataset.
The results are shown in Table \ref{edgeexp1}.
It can be seen that the changes in parameter $\alpha$ in this range have little impact on the performance.
The existence of edge loss in loss functions can already improve the network's attention on the details such as boundary regions and small areas.}

\editc{Furthermore, we use the edge loss to train the existing SOTA methods.
The results are shown in Table \ref{edgeexp2}.
It can be seen that our proposed edge loss is also helpful for existing methods.}

\begin{table}[ht]
\caption{\editc{Experiment with Different $\alpha$ in Edge Loss}}
\label{edgeexp1}
\centering
\begin{tabular}{cccccc}
\hline
\hline
$\alpha$ & 0.1 & 0.5 & 1 & 3 & 5\\
\hline
Precision & 0.962 & 0.964 & 0.962 & 0.964 & 0.963\\
Recall & 0.970 & 0.970 & 0.971 & 0.971 & 0.972\\
F1 & 0.966 & 0.967 & 0.967 & 0.967 & 0.967\\
\hline
\hline
\end{tabular}
\end{table}

\begin{table}[ht]
\caption{\editc{Experiment with Edge Loss combined with different methods}}
\label{edgeexp2}
\centering
\begin{tabular}{ccccc}
\hline
\hline
Method & Edge Loss & Precision & Recall & F1\\
\hline
\multirow{2}*{FC-Siam-diff} & - & 0.774 & 0.631 & 0.688\\
& \checkmark & 0.802 & 0.629 & 0.692\\
\hline
\multirow{2}*{UNet++\_MSOF} & - & 0.946 & 0.939 & 0.943\\
& \checkmark & 0.960 & 0.966 & 0.960\\
\hline
\multirow{2}*{SNUNet-CD} & - & 0.961 & 0.965 & 0.963\\
& \checkmark & 0.964 & 0.970 & 0.967\\
\hline
\hline
\end{tabular}
\end{table}

\subsection{Experiment in Efficient Training Strategy}

\begin{figure*}[ht]
\centering
\subfloat[]{
    \includegraphics[width= 0.23\linewidth]{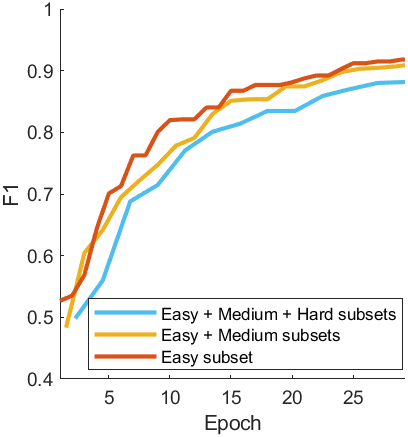}
}
\subfloat[]{
    \includegraphics[width= 0.23\linewidth]{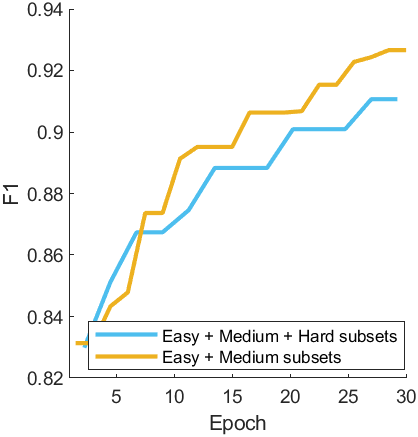}
}
\subfloat[]{
    \includegraphics[width= 0.23\linewidth]{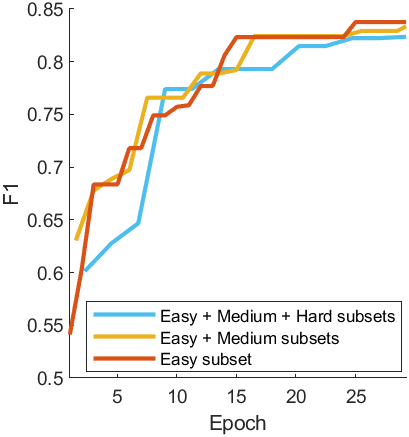}
}
\subfloat[]{
    \includegraphics[width= 0.23\linewidth]{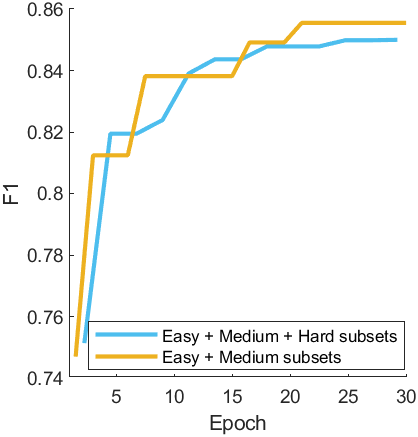}
}
\caption{Epoch versus F1.
(a) and (b) are on CDD dataset.
(c) and (d) are on LEVIR-CD dataset.
The blue line indicates that the easy, medium and hard subsets are used for training.
The orange line indicates that the easy and medium subsets are used for training.
The red line indicates that only the easy subset is used for training.
(a) and (c) are in the early stage of training.
(b) and (d) are in the middle stage of training.}
\label{hierexp}
\end{figure*}

To test the efficient training strategy further, we trained the network with datasets of different difficulties.
Fig. \ref{hierexp} shows epoch versus F1 on the validation set.
Fig. \ref{hierexp}(a) and (c) indicate that different subsets are input for training in the early stage of training.
It can be seen that the red line rises the fastest and achieves the highest F1, while the blue line performs worst.
It can be proved that in the early stage of network training, feeding easy tasks to the network can not only reduce the amount of computation and speed up the network operation, but also achieve better training results.
Fig. \ref{hierexp}(b) and (d) indicate that different subsets are input for training after some training epochs with the easy subset.
It can be seen that the orange line rises faster and achieves higher F1, while the blue line performs worse.
It can be proved that in the middle stage of network training, feeding moderate tasks to the network can reduce the amount of computation, speed up the network operation, and achieve better training results.
These couple of experiments further prove the effectiveness of our proposed efficient training strategy.

\begin{table}[ht]
\caption{\editb{Experiment between Efficient Training Strategy and Random Sampling Strategy}}
\label{Adaptive1}
\centering
\begin{tabular}{ccccc}
\hline
\hline
Dataset & Method & Precision & Recall & F1\\
\hline
\multirow{2}*{CDD} & Efficient Training Strategy & 0.967 & 0.977 & 0.972\\
 & Random Sampling Strategy & 0.962 & 0.973 & 0.967\\
\hline
\multirow{2}*{LEVIR-CD} & Efficient Training Strategy & 0.915 & 0.888 & 0.901\\
 & Random Sampling Strategy & 0.888 & 0.873 & 0.878\\
\hline
\hline
\end{tabular}
\end{table}


We also compare our efficient training strategy with the random sampling strategy.
The probability of each sample is set to $e^{-L}$, where $L$ represents the detection loss of the initial model.
Three-quarters of the samples are randomly selected for each training epoch, according to their probabilities.
The results are shown in Table \ref{Adaptive1}.
It can be proved that our efficient training strategy can achieve better results compared with the random sampling strategy.

\begin{table*}[ht]
\caption{\editb{Experiment with Efficient Training Strategy combined with different methods}}
\label{efficient1}
\centering
\begin{tabular}{cccccc}
\hline
\hline
Dataset & Method & Efficient Training & Precision & Recall & F1\\
\hline
\multirow{6}*{CDD} & \multirow{2}*{FC-Siam-diff} & - & 0.774 & 0.631 & 0.688\\
& & \checkmark & 0.847 & 0.607 & 0.697\\
\cline{2-6}
& \multirow{2}*{UNet++\_MSOF} & - & 0.946 & 0.939 & 0.943\\
& & \checkmark & 0.959 & 0.963 & 0.961\\
\cline{2-6}
& \multirow{2}*{SNUNet-CD} & - & 0.961 & 0.965 & 0.963\\
& & \checkmark & 0.965 & 0.973 & 0.969\\
\hline
\multirow{6}*{LEVIR-CD} & \multirow{2}*{FC-Siam-diff} & - & 0.883 & 0.785 & 0.826\\
& & \checkmark & 0.878 & 0.804 & 0.834\\
\cline{2-6}
& \multirow{2}*{UNet++\_MSOF} & - & 0.906 & 0.886 & 0.894\\
& & \checkmark & 0.923 & 0.878 & 0.899\\
\cline{2-6}
& \multirow{2}*{SNUNet-CD} & - & 0.907 & 0.887 & 0.896\\
& & \checkmark & 0.908 & 0.892 & 0.900\\
\hline
\hline
\end{tabular}
\end{table*}


Furthermore, we use the efficient training strategy to train the existing SOTA methods.
The results are shown in Table \ref{efficient1}.
It can be seen that our proposed efficient training strategy is also helpful for existing methods.
The CNN could learn more powerful features from easy to hard with fewer FLOPs (about 86.5\%) and achieve better performance.

\section{Conclusion}

In this paper, we propose RDP-Net, an effective method for change detection in remote sensing.
It contains an efficient training strategy, an effective edge loss and a brand new CNN backbone focusing on preserving regions' details.
The efficient training strategy splits the dataset by difficulty level, allows the network to learn from easy to hard and achieve a better performance compared with reducing the probability of hard samples in the training process.
%
Edge loss increases the penalty for errors on the boundary regions when calculating loss, which can improve the network's attention on the details such as boundary regions and small areas, and enhance the detection performance.
%
The proposed RDP-Net achieves the state-of-the-art empirical performance with only 1.70M parameters.
The F1-score is 3.6\% higher than that of a comparable size network (3.31M parameters).
The number of parameters is only 12.9\% of SOTA (13.21M parameters).
We performed experiments on CDD and LEVIR-CD datasets.
The experimental results show that the proposed change detection method can significantly improve the accuracy of the task of remote sensing change detection,
and the efficient training strategy can also benefit other methods.



 
\bibliography{references}
%

\bibliographystyle{IEEEtran}

\vfill

\end{document}